\documentclass{aastex631}
\usepackage{arydshln}
\usepackage{booktabs}
\usepackage{multirow}
\usepackage{subfigure}
\usepackage{multirow}
\usepackage{epstopdf}
\usepackage[normalem]{ulem}
\usepackage[figuresright]{rotating}
\useunder{\uline}{\ul}{}

\newcommand{\velunit}{km~s$^{-1}$}

\shorttitle{}
\shortauthors{Shi et al.}
\graphicspath{{./}{figures/}}

\begin{document}

\title{The differences in the origination and properties of the near-Earth solar wind between solar cycles 23 and 24}

\correspondingauthor{Hui Fu}
\email{fuhui@sdu.edu.cn}

\author{Xinzheng Shi}
\affiliation{Shandong Key Laboratory of Optical Astronomy and Solar-Terrestrial Environment, Institute of Space Sciences, Shandong University, Weihai, Shandong, 264209, China}

\author{Hui Fu}
\affiliation{Shandong Key Laboratory of Optical Astronomy and Solar-Terrestrial Environment, Institute of Space Sciences, Shandong University, Weihai, Shandong, 264209, China}

\author{Zhenghua Huang}
\affiliation{Shandong Key Laboratory of Optical Astronomy and Solar-Terrestrial Environment, Institute of Space Sciences, Shandong University, Weihai, Shandong, 264209, China}

\author{Limei Yan}
\affiliation{Key Laboratory of Earth and Planetary Physics, Institute of Geology and Geophysics, Chinese Academy of Sciences, Beijing, 100029, China}

\author{Chi Ma}
\affiliation{Shandong Key Laboratory of Optical Astronomy and Solar-Terrestrial Environment, Institute of Space Sciences, Shandong University, Weihai, Shandong, 264209, China}

\author{Chenxi Huangfu}
\affiliation{Shandong Key Laboratory of Optical Astronomy and Solar-Terrestrial Environment, Institute of Space Sciences, Shandong University, Weihai, Shandong, 264209, China}

\author{Hongqiang Song}
\affiliation{Shandong Key Laboratory of Optical Astronomy and Solar-Terrestrial Environment, Institute of Space Sciences, Shandong University, Weihai, Shandong, 264209, China}

\author{Lidong Xia}
\affiliation{Shandong Key Laboratory of Optical Astronomy and Solar-Terrestrial Environment, Institute of Space Sciences, Shandong University, Weihai, Shandong, 264209, China}




\begin{abstract}
The dependence of the sources and properties of the near-Earth solar wind on solar cycle activity is an important issue in solar and space physics. We use the improved "two-step" mapping procedure that takes into account the initial acceleration processes to trace the near-Earth solar winds back to their source regions from 1999 to 2020, covering solar cycles (SCs) 23 and 24. Then the solar wind is categorized into coronal hole (CH), active region (AR), and quiet Sun (QS) solar wind based on the source region types.
We find that the proportions of CH and AR (QS) wind during SC 23 are higher (lower) than those during SC 24. During solar maximum and declining phases, the magnetic field strength, speed, helium abundance ({$A_{He}$}), and charge states of all three types of solar wind during SC 23 are generally higher than those during SC 24. During solar minimum, these parameters of solar wind are generally lower during SC 23 than those during SC 24. There is a significant decrease in the charge states of all three types of solar wind during the solar minimum of SC 23. The present statistical results demonstrate that the sources and properties of the solar wind are both influenced by solar cycle amplitude. The temperatures of AR, QS, and CH regions exhibit significant difference at low altitudes, whereas they are almost uniform at high altitudes.

\end{abstract}

\keywords{Solar wind (1534); Solar cycle (1487); Solar activity (1475); Solar abundances (1474)}

\section{Introduction}\label{sec:intro}
Sunspot numbers (SSNs) serve as a reliable indicator of solar activity \citep{2000Natur.408..445S,2009JGRA..114.9105G,2013JGRA..118.7525G}. The about 11-year cycle of SSNs was discovered by Samuel Heinrich Schwabe in 1844 \citep{1844AN.....21..233S}. The statistical results show that the SSNs change significantly from solar maximum to minimum, and they are also not the same in different solar cycles \citep{2015LRSP...12....4H}.

The properties of different layers of the solar atmosphere are closely connected to SSNs \citep{1952BAICz...3...52K,1979Natur.280...24D,1994SoPh..152...53A,2008ApJ...686L..41R,2013SSRv..176..237F,2014Ap&SS.350..479R}. The SSN is a good indicator of the magnetic field strength of the photosphere and it is proportional to the sunspot areas \citep{2002A&A...390..707T,2003SoPh..215..111T}. The total solar irradiance (TSI) is predominantly contributed by the radiation from the photosphere and it also presents a solar cycle variation although it is only 0.1\% higher during solar maximum than during solar minimum \citep{2003ESASP.535..183F,2005AdSpR..35..376S}. The TSI is extremely lower during the solar minimum between SCs 23 and 24 compared to the previous solar minimums \citep{2013SSRv..176..237F}.

The Ly-a irradiance, Ca II K, and Mg II indices which represent the activities of the chromosphere are also proportional to the SSNs \citep{1986JGR....91.8672H,2000JGR...10527195W,2013SSRv..176..237F}. The intensity of the corona which is the outermost layer of the solar atmosphere is also positively correlated with the SSNs. The intensity of the coronal green line Fe XIV 5303 \AA\, is higher (lower) during solar maximum (minimum) \citep{2005JGRA..110.8106R,2014Ap&SS.350..479R}. The background X-ray flux during solar maximum is 3 orders of magnitude higher than that during solar minimum \citep{1994SoPh..152...53A}. In addition, the heliospheric magnetic field (HMF) is also modulated by the SSNs and it is higher (lower) during solar maximum (minimum) \citep{2000Natur.408..445S,2003JGRA..108.1128L,2008GeoRL..3520108O,2013JGRA..118.7525G,2017ApJ...837..165R}.

The activities, such as flares and coronal mass ejections (CMEs) are also closely related to the SSNs. The flares and CMEs are the two most powerful energy release processes on the Sun. The flare-index \citep{1952BAICz...3...52K}, which reflects the total energy released by flares, is positively correlated with the SSNs \citep{2003SoPh..214..375O,2004SoPh..223..287O}. The occurrence rate of CMEs is also changed with the SSNs and it is much higher during solar maximum than that during solar minimum \citep{2001ApJ...556..432C,2010SoPh..264..189R,2016SoPh..291.2419C,2022ApJ...940..103S}. In addition, \cite{2022ApJ...940..103S} divided the ICMEs detected near the Earth into two types: with and without flares, only based on the charge states inside ICMEs. They found that not only the yearly number of ICMEs associated with flares is changed with SSNs, but the occurrence rate of ICMEs without flares is also positively correlated with SSNs.

The source of the solar wind also depends on the solar activity phases. The large-scale magnetic field structure of the Sun is approximated as a dipole field during solar minimum. The high and middle latitude area is occupied by large polar coronal holes. In contrast, the low latitude is covered by stable streamers. Therefore, the high and middle latitude is filled by the fast solar wind originating from large polar coronal holes. Whereas, low latitude is occupied by the slow solar wind coming from streamers and/or coronal hole boundaries \citep{1996RvGeo..34..379G,2006SSRv..124...51S,2008GeoRL..3518103M,2009ApJ...691..760W,2016SSRv..201...55A}. On the other hand, the magnetic field structures are more complex during solar maximum. Thus the solar wind distribution characteristics in the heliosphere are also complicated as the large polar coronal holes are dismissed during solar maximum. \cite{2015SoPh..290.1399F} traced the near-Earth solar wind back to the Sun using the standard two-step mapping procedure. Then the footpoint regions of the solar wind were classified into coronal hole (CH), active (AR), and quiet Sun (QS) regions based on the photospheric magnetic field and corona structures. Finally, the solar wind detected near the Earth was categorized into three types by the associated source region types. They found that about half of the near-Earth solar wind comes from the active regions during solar maximum. The proportion of CH wind is the highest during the solar declining phase. Whereas, nearly half of the near-Earth solar wind originates from quiet Sun regions during solar minimum.

The in-situ parameters of the solar wind are closely related to the SSNs as well. The charge states and elemental abundance in the solar wind are higher (lower) during solar maximum (minimum). The charge states, such as $O^{7+}/O^{6+}$, do not change beyond several solar radii. Therefore, it is believed that the $O^{7+}/O^{6+}$ inside solar wind reflects the temperature of source regions \citep{1983ApJ...275..354O,1986SoPh..103..347B,2000JGR...10510527H,2012ApJ...758L..21L}. The statistical results show that the $O^{7+}/O^{6+}$ of near-Earth solar wind is decreased from solar maximum to minimum \citep{2015SoPh..290.1399F,2017ApJ...836..169F,2022ApJ...940..103S}. The above results are consistent with the belief that the coronal temperature is highest (lowest) during solar maximum (minimum). The solar cycle variations of helium abundance ($A_{He}$) inside the fast and slow solar wind are significantly different. The $A_{He}$ of fast solar wind is higher (about 4-6) with a small distribution range and does not change with solar activity. Whereas the $A_{He}$ inside the slow solar wind is lower with a large distribution range and it is positively correlated with solar activity \citep{2001GeoRL..28.2767A,2007ApJ...660..901K,2012ApJ...745..162K,2018MNRAS.478.1884F,2019ApJ...879L...6A}. The FIP bias is higher (lower) inside slow (fast) solar wind with a large (small) distribution range and it is also proportional to SSNs \citep{2017ApJ...836..169F}. In addition, the charge states, $A_{He}$, and FIP bias inside the solar wind coming from coronal hole, active, and quiet Sun regions are all decreased from solar maximum to minimum \citep{2015SoPh..290.1399F,2017ApJ...836..169F,2018MNRAS.478.1884F}.

The previous studies demonstrate that the occurrence rate and properties of ICMEs are significantly influenced by solar cycle amplitude. The SSNs of SC 24 are significantly lower than that of SC 23. The yearly SSNs are nearly 180 during the SC 23 maximum. In contrast, the yearly SSNs are no more than 120 during the SC 24 maximum. Accordingly, the occurrence and properties of ICMEs are different between SCs 23 and 24. \cite{2020JPhCS1620a2005G} found that the occurrence, spatial scale, and speed of ICMEs are all different during the SCs 23 and 24. The occurrence rate of CMEs observed by LASCO/SOHO is higher during SC 24 than SC 23. However, the number of fast and wide CMEs during SC 23 is higher than that of SC 24. They suggested that this is consistent with the fact that the solar cycle amplitude is higher during SC 23 than that of SC 24. \cite{2022ApJ...925..137S} analyzed the helium abundance inside ICMEs and solar wind from 1998 to 2019. They found that the $A_{He}$ inside ICMEs is proportional to the SSNs and the $A_{He}$ inside ICMEs during the SC 23 is higher than that of ICMEs during the SC 24 (see Figure 2 in \cite{2022ApJ...925..137S}). In addition, \cite{2022ApJ...940..103S} found that the durations, speed, charge states ($Q_{Fe}$ and $O^{7+}/O^{6+}$), helium abundance, and FIP bias are all lower for ICMEs without flares on the Sun (non-flare-CMEs/NFCs) than that for ICMEs with flares on the Sun (flare-CMEs/FCs). The occurrence rate of FCs is much higher during SC 23 than that during SC 24. The yearly number of FCs during SC 23 is about three times that of SC 24 \citep{2022ApJ...940..103S}.

Are the sources and properties of the near-Earth solar wind influenced by the solar cycle amplitude? Whether the proportions of the solar wind originating from different source regions are the same or not between the solar cycles with diverse SSNs? Are there differences in properties of the solar wind originating from the same type of source regions between different solar cycles? The fact that the SSNs during the SC 24 are significantly lower than those of the SC 23 allows us to examine the above questions. The photospheric magnetic field, structures of the corona, and in-situ properties of the near-Earth solar wind are observed and measured continuously during SCs 23 and 24. Consequently, we can compare the sources and properties of the near-Earth solar wind between SCs 23 and 24.

In the present study, we first attempt to improve the standard two-step mapping procedure by taking into account the two effects, the interactions between the fast and slow solar wind streams and the initial acceleration process of the solar wind. In addition, the improved "two-step" procedure is tested by statistical analysis. Then the solar wind detected by the ACE and Wind satellites is traced back to the solar surface. The sources of the solar wind are classified into CH, AR, and QS regions based on the photospheric magnetic field and the coronal structures \citep{2015SoPh..290.1399F,2017ApJ...836..169F,2018MNRAS.478.1884F}. Accordingly, the solar wind near the Earth is categorized by the types of source regions. Finally, the sources and properties of the three types of solar wind are analyzed and compared between SCs 23 and 24. The statistical results will provide more observational facts on the sources and properties of the near-Earth solar wind. In addition, the outcome will improve our understanding of the influence of solar cycle amplitude on the sources and properties of the solar wind.

The paper is organized as follows. The data and the improved "two-step" mapping procedure are described and introduced in Section 2. The statistical results of the solar wind during SCs 23 and 24 are presented and discussed in Section 3. In section 4, the summary of the results and the concluding remarks are presented.

\section{Instruments and data analysis} \label{sec:style}
The in-situ parameters of the solar wind in the present study are obtained from the Wind\footnote{\url{https://cdaweb.gsfc.nasa.gov/index.html}} \citep{1995SSRv...71....5A} and Advanced Composition Explorer \citep[ACE\footnote{\url{http://www.srl.caltech.edu/ACE}},][]{1998SSRv...86....1S} spacecraft. The Solar Wind Experiment \citep[SWE,][]{1995SSRv...71...55O} onboard Wind provides data on speed and helium abundance ($A_{He}$), and the Magnetic Field Investigation \citep[MFI,][]{1995SSRv...71..207L} onboard Wind measures the interplanetary magnetic field. The Solar Wind Ion Composition Spectrometer \citep[SWICS,][]{1998SSRv...86..497G} onboard ACE provides the charge states ($Q_{Fe}$, $O^{7+}/O^{6+}$) of the solar wind.


The larger magnetic field structure of the Sun is extrapolated by the PFSS model. The magnetic field vectors are provided by the Solar Software PFSS package \citep{2003SoPh..212..165S}. The photospheric magnetograms measured by MDI/SOHO \citep{1995SoPh..162..129S} and HMI/SDO \citep[from 2011 to 2020,][]{2012SoPh..275..229S} are used as the bottom boundary. The coronal structures are observed by EIT/SOHO \citep[from 1999 to 2010,][]{1995SoPh..162..291D} and AIA/SDO \citep[from 2011 to 2020,][]{2012SoPh..275...17L}. The synoptic images of solar corona (EIT\footnote{\url{http://satdat.oulu.fi/solar_data/Synoptic_Map_EIT_AIA/EIT_AIA_Synop_Maps/EIT/195A/195A_Lat/195A_Lat_fits/}}, AIA\footnote{\url{https://sdo.gsfc.nasa.gov/data/synoptic/}}) are used in the present study. The yearly SSNs are provided by the Solar Influence Data Center of the Royal Observatory of Belgium\footnote{\url{http://www.sidc.be/silso/home}}.

In the present study, the standard two-step mapping procedure for tracing the solar wind back to the Sun is improved. Generally, the standard two-step mapping procedure is used in connecting the solar winds detected in situ and their source regions \citep{1998JGR...10314587N,2004SoPh..223..209L,2015SoPh..290.1399F,2017ApJ...836..169F,2018MNRAS.478.1884F,2017ApJ...846..135Z}. The solar wind detected in situ is first traced back to the source surface at 2.5 R$_{\bigodot}$ assuming the solar wind propagates with a constant speed. Then the wind parcel is traced back to the solar surface following the magnetic field lines derived by the potential-field-source-surface (PFSS) model \citep{1969SoPh....6..442S,1969SoPh....9..131A}. However, it's evident that the solar wind speed does not remain constant during the propagation in the heliosphere. The two processes during the propagation of the solar wind in the heliosphere should not be ignored. The first effect is the initial acceleration process of the solar wind. Generally, the fast solar wind is fully accelerated at about 10 R$_{\bigodot}$ \citep{2007ApJS..171..520C,2009LRSP....6....3C}. In contrast, the acceleration process of the slow solar wind finished farther compared with that of the fast solar wind. The theoretical models demonstrate that the acceleration of the slow solar wind should be finished at about 20 R$_{\bigodot}$ \citep{2009LRSP....6....3C,2016SSRv..201...55A}. The observational results about the speed curve of blobs are also consistent with the notion that the acceleration process of slow solar wind is finished at about 20 R$_{\bigodot}$ \citep{1997ApJ...484..472S,2009SoPh..258..129S,2012SoPh..276..261S}. The second effect is the interaction between the slow and fast solar wind streams during the propagation in the heliosphere. The slow solar wind should be accelerated if it is followed by a fast stream. For the same reason, the fast solar wind should be decelerated if it is behind a slow stream.

In the present study, the two improvements are considered during the tracing procedure.
First, the initial acceleration processes of the solar wind are included. The solar wind is first classified into fast and slow solar wind with a threshold of 500 \velunit. We make a simple treatment for correcting the initial acceleration processes of the solar wind. We assume that the solar wind accelerates uniformly below 10 (20) R$_{\bigodot}$ for fast (slow) solar wind. Therefore, the average speed for the fast (slow) solar wind below 10 (20) R$_{\bigodot}$ is regarded as half of the near-Earth speed. Second, the effects of the interaction between fast and slow streams during propagation in the heliosphere are also evaluated. Qualitatively, the fast (slow) solar wind should be decelerated (accelerated) if it is behind (followed by) a slow (fast) stream. Following the method adopted by \cite{1997ApJ...488L..51W}, we introduce a modification factor (MF) for correcting the solar wind stream interactions. The speed of behind fast (ahead slow) solar wind will be modified to detected\_v +MF*delta\_v (detected\_v -MF*delta\_v). Here, the detected\_v is the speed of the solar wind detected near the Earth and the delta\_v is the speed difference between the two solar wind parcels.

The important way to evaluate the accuracy of the tracing procedures is to examine the polarity consistency between the in-situ solar wind and its footpoint region. To choose the best mapping procedure, we establish four experimental groups. First, use the standard two-step mapping procedure as the control group. Second, consider only the impacts of the initial acceleration process. Third, consider only the effects of the interaction between fast and slow streams during propagation in the heliosphere. Fourth, simultaneously consider both of the above two effects. In the present study, the time resolution for the near-Earth solar wind traced back to the Sun is 12 hours.

The statistical analyses demonstrate that the initial acceleration processes should be included in the tracing back procedures. Using the standard two-step mapping procedure, the overall matched polarity proportion is 81.53\%, from 1999 to 2020. When we only make corrections to the initial acceleration processes, the matched polarity proportion increases to 82.38\%, resulting in an improvement of 0.85\% compared to the standard two-step mapping procedure. However, when considering the effects of the interaction between fast and slow streams during propagation with different MF, the increase of the overall matched polarity proportion does not exceed 0.2\%. For some MFs, it may even have negative impacts. Furthermore, when considering the combined effects of these two processes, the improvement in matched polarity proportion is similar to the procedure for only including the effects of the initial acceleration processes. Therefore, the impact of the interaction between fast and slow streams may not be as significant as initially expected. Hence, in subsequent analyses, we will focus solely on correcting the effects of the initial acceleration process.

\begin{figure}[!h]
\centering
\includegraphics[width=1.0\textwidth]{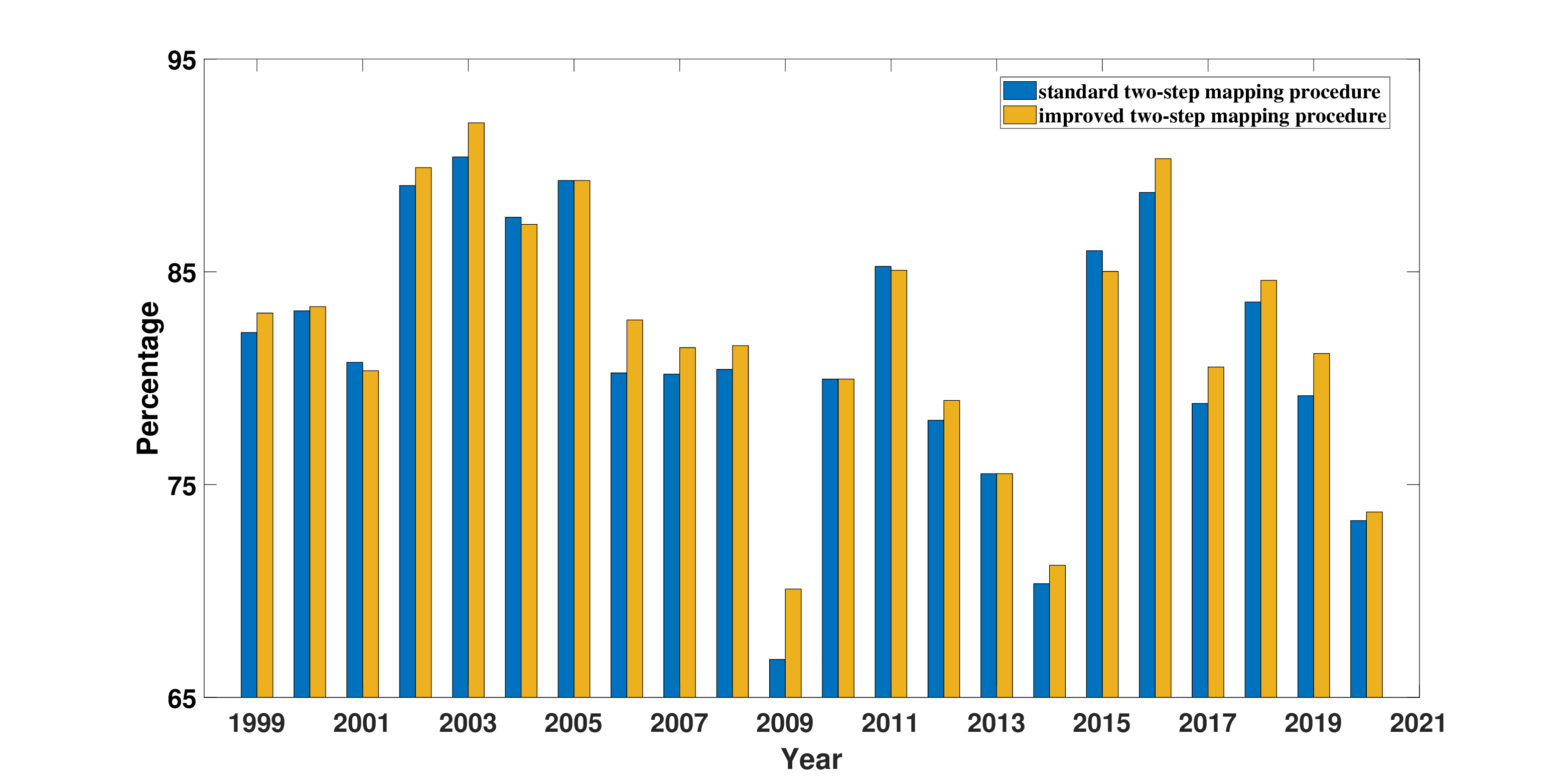}
\caption{The yearly matched polarity proportions between footpoint regions and the in-situ solar wind for the standard (blue) and improved (yellow) two-step mapping procedures. Statistically, the matched polarity proportions of the improved procedure are higher than those of the standard two-step mapping procedure.}
\label{mpp}
\end{figure}

The yearly matched polarity proportions between footpoint regions and the in-situ solar wind for the standard and improved two-step mapping procedures are presented in Figure \ref{mpp}. The yearly matched polarity proportions are highest during the solar declining phase for both standard and improved two-step mapping procedures. The above characteristic is the same as that in \cite{2015SoPh..290.1399F}. Statistically, the matched polarity proportions of the improved two-step mapping procedure are higher than those of the standard two-step mapping procedure. The matched polarity proportions for the improved procedures are higher than or equal to those of standard procedures in 18 years from 1999 to 2020. The above statistical results suggest that the improved two-step mapping procedure can trace the solar wind back to its source region more credibly.

\begin{figure}[!h]
\centering
\includegraphics[width=0.8\textwidth]{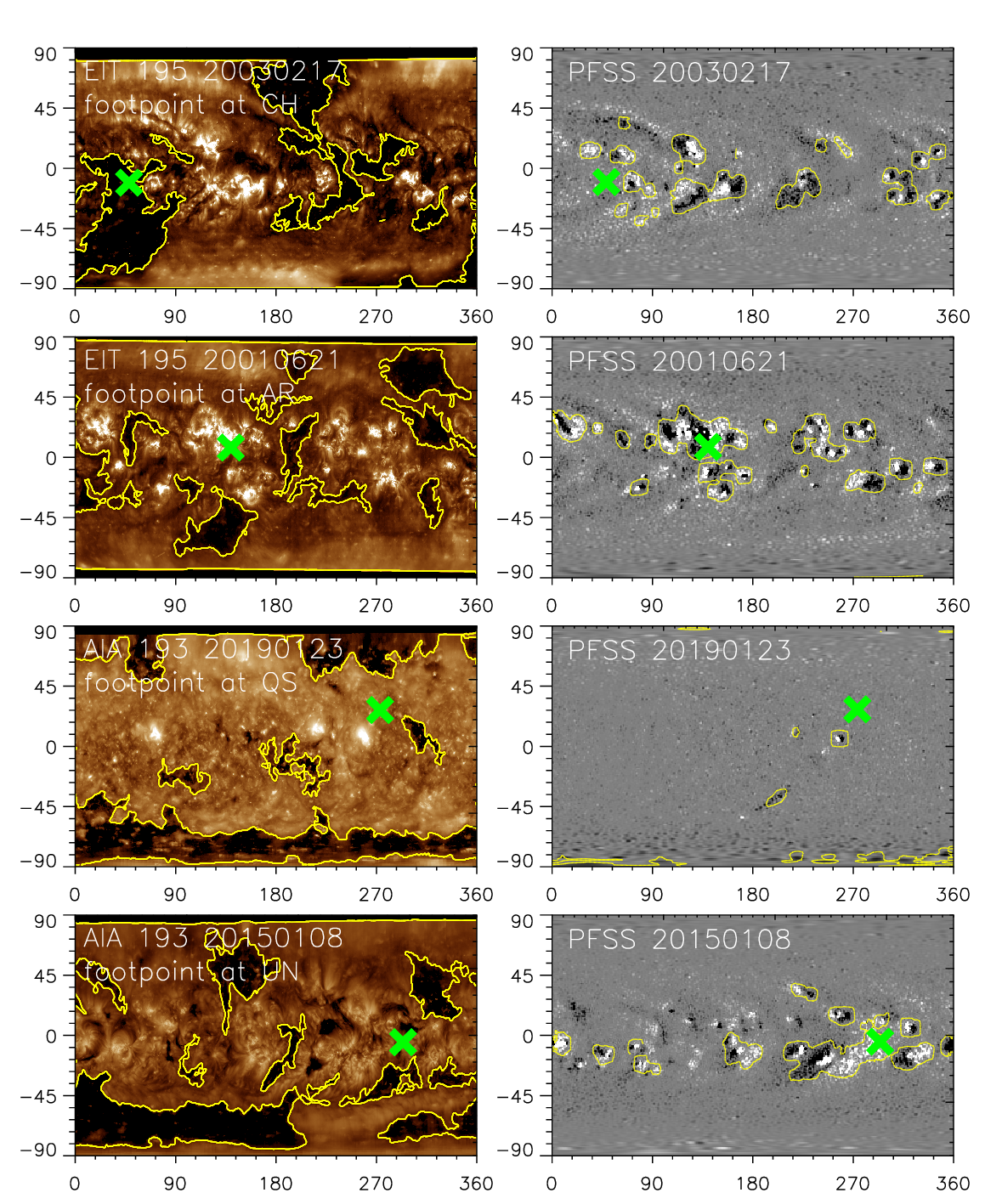}
\caption{Illustration of the classification scheme of the solar wind based on the source region types. The footpoints of the solar wind are denoted by green crosses. According to the source region types, the solar wind is categorized into CH (first row), AR (second row), QS (third row), and UN (fourth row) solar wind. The EUV synoptic images are shown on the left column, and the corresponding photospheric magnetic field images are shown on the right column. The yellow outlines denote the boundaries of coronal holes and magnetically concentrated areas.}
\label{footpoint}
\end{figure}

Then the solar wind is classified into three categories by the source region types. The classification scheme is demonstrated in Figure \ref{footpoint}. First, the footpoints of the near-Earth solar wind are overplotted on the photospheric magnetic field and EUV synoptic images. Second, the solar wind is classified into CH, AR, and QS solar wind based on the footpoint locations. The classification relies on the boundaries of coronal holes and magnetically concentrated areas. The solar wind is categorized into CH wind if its footpoint is located inside the boundaries of coronal holes. The AR wind is that the footpoint is located inside the magnetically concentrated areas with NOAA AR number. The QS wind is that the footpoint is outside the coronal holes and magnetically concentrated areas. There are also some footpoints located inside magnetic field concentrated areas that are not numbered by NOAA, and these are defined as undefined (UN). More details on the determination of boundaries and classification procedures can be found in \cite{2015SoPh..290.1399F,2017ApJ...836..169F,2018MNRAS.478.1884F}.

The intervals occupied by ICMEs are removed by the near-Earth ICMEs list organized by Richardson and Cane \citep[RC list\footnote{\url{https://izw1.caltech.edu/ACE/ASC/DATA/level3/icmetable2.htm}},][]{2003JGRA..108.1156C,2004JGRA..109.9104R,2010SoPh..264..189R}, as we concentrated on the steady-state solar wind. The data analyzed in the present study covers the years 1999 to 2020. The period includes SCs 23 and 24, and the hourly solar wind samples for the solar wind as a whole, CH, AR, QS, and UN wind are 120216, 45348, 31992, 24384, and 18492, respectively.

\section{Results and discission} \label{sec:floats}

\begin{table}[htbp]
  \centering
  \caption{Numbers of hourly solar wind and ICMEs analyzed in each year}
    \begin{tabular}{cccccc}
    \midrule
    \midrule
    Year  & CH winds & AR winds & QS winds & Undefined & ICMEs \\
    1999  & 1092  & 2472  & 324   & 816   & 1014 \\
    2000  & 1824  & 2376  & 216   & 684   & 1561 \\
    2001  & 1200  & 2388  & 336   & 984   & 1534 \\
    2002  & 2412  & 2604  & 372   & 1020  & 853 \\
    2003  & 3312  & 2112  & 600   & 864   & 612 \\
    2004  & 2364  & 1632  & 144   & 1212  & 666 \\
    2005  & 2520  & 1392  & 636   & 780   & 843 \\
    2006  & 1968  & 948   & 1320  & 624   & 294 \\
    2007  & 2580  & 732   & 1980  & 348   & 34 \\
    2008  & 2436  & 636   & 2772  & 180   & 65 \\
    2009  & 1860  & 96    & 2484  & 132   & 228 \\
    2010  & 1428  & 972   & 1392  & 1344  & 370 \\
    2011  & 1464  & 2580  & 504   & 1128  & 641 \\
    2012  & 1572  & 2220  & 468   & 828   & 867 \\
    2013  & 1164  & 1836  & 372   & 1356  & 644 \\
    2014  & 900   & 1608  & 216   & 1224  & 566 \\
    2015  & 2016  & 1800  & 588   & 1860  & 703 \\
    2016  & 2532  & 1188  & 1320  & 1176  & 358 \\
    2017  & 2832  & 984   & 1332  & 1056  & 276 \\
    2018  & 3444  & 288   & 2136  & 264   & 189 \\
    2019  & 2292  & 240   & 2916  & 240   & 224 \\
    2020  & 2136  & 888   & 1956  & 372   & 117 \\
    \midrule
    \end{tabular}%
  \label{tab:label}%
\end{table}%

\subsection{The comparison of the sources of the near-Earth solar wind between solar cycles 23 and 24}
The numbers of hourly samples of CH (second column), AR (third column), QS (fourth column), and UN (fifth column) wind for the years of 1999-2020 are presented in Table \ref{tab:label}. For completeness, the periods occupied by ICMEs are also presented in the sixth column. Only the solar wind that the in-situ magnetic polarity is consistent with the footpoint is analyzed in the present study. The 12-hour solar wind parcels with the average in-situ magnetic field strength less than 0.5 nT in the Sun-Earth direction are removed, as they should come from different source regions with different magnetic polarities.
In total, 59784 (60432) hourly samples of the solar wind are analyzed during the SC 23 (24). The samples of CH, AR, QS, and UN wind are 23568 (21780), 17388 (14604), 11184 (13200), and 7644 (10848) during SC 23 (24).

\begin{figure}[!h]
\centering
\includegraphics[width=1.0\textwidth]{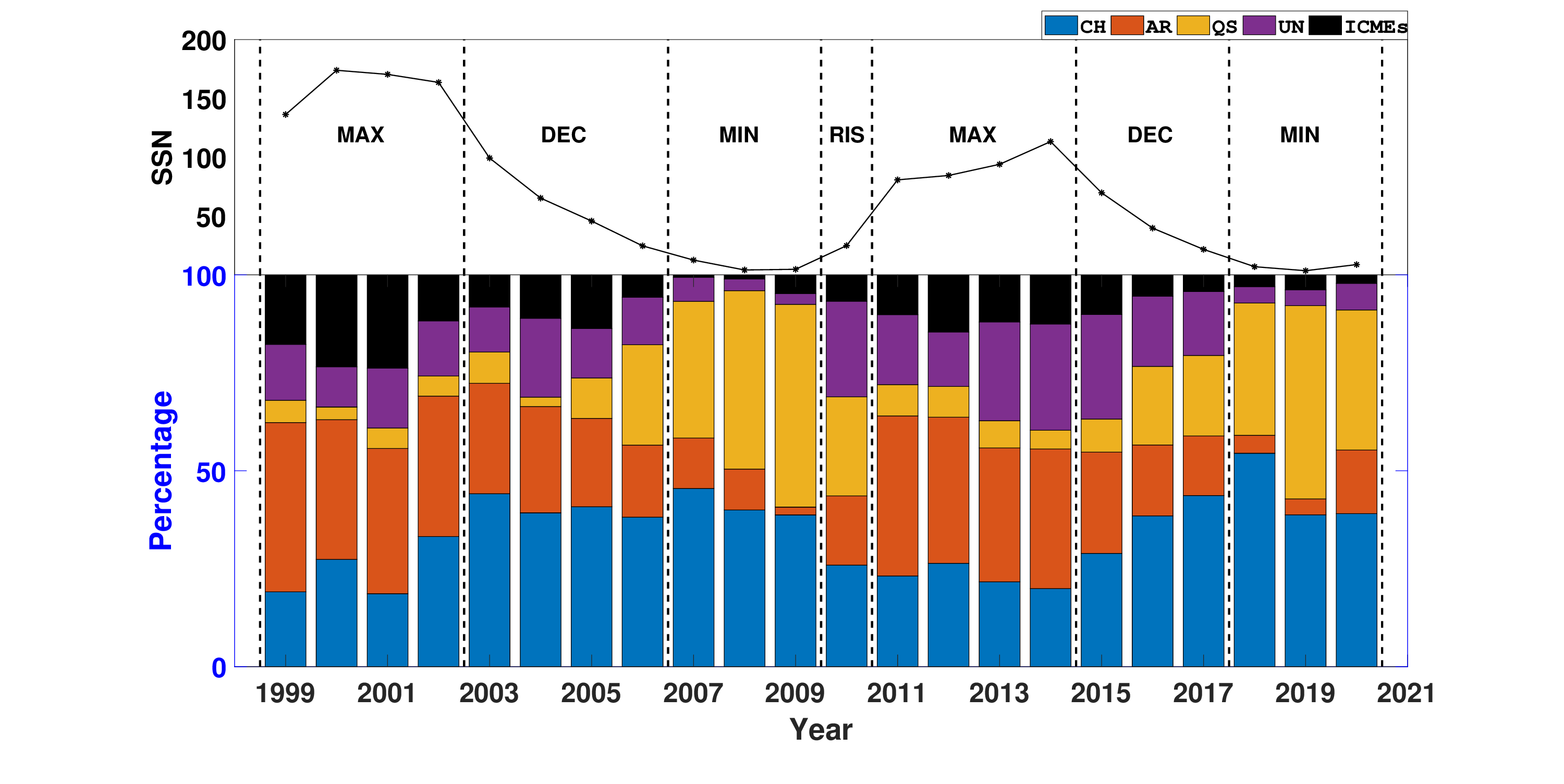}
\caption{The sunspot numbers (SSNs) and the proportions of different types of solar wind in each year during 1999-2020. The period is divided into different solar activity phases based on the SSNs (top panel). The CH, AR, QS, and UN steady solar wind are denoted by blue, red, yellow, and purple, respectively (bottom panel). The proportions of near-Earth ICMEs are also included, and it is represented by black. The proportions of AR (QS) wind and ICMEs are significantly decreased (increased) with the decrease of solar activity.}
\label{year}
\end{figure}

The proportions of ICMEs and different types of the near-Earth solar wind in each year during 1999-2020 are presented in the bottom panel of Figure \ref{year}. The SSNs in each year during 1999-2020 are given in the top panel of Figure \ref{year}. The SCs 23 and 24 are further divided into different solar activity phases based on the SSNs, as the sources and properties of solar wind all change with solar cycle phases \citep{2009GeoRL..3614104Z,2015SoPh..290.1399F,2017ApJ...836..169F,2018MNRAS.478.1884F}. The solar maximum, declining, and minimum phases during SCs 23 and 24 are analyzed and compared in the present study. The maximum phase of SC 23 (24) spans from 1999-2002 (2011-2014), the declining phase of SC 23 (24) spans from 2003-2006 (2015-2017), and the minimum phase of SC 23 (24) spans from 2007-2009 (2018-2020). It should be noted that the rising phase interval is not included in the statistical analysis because of the small sample of data available when compared to the other solar activity phases. The CH, AR, QS, and UN steady solar wind are denoted by blue, red, yellow, and purple, respectively. The proportions of different types of solar wind are all changed with solar activity phases. The proportions of AR (QS) wind are significantly decreased (increased) with the decrease of solar activity. The proportion of CH wind during solar maximum is lower compared with those during solar declining and minimum phases (see blue bars in the bottom panel of Figure \ref{year}). The periods occupied by ICMEs (black bars) are also significantly decreased with a decrease of SSNs. The statistical results are consistent with the fact that the number of active regions decreases with solar activity. The above characteristics are the same as those reported by \cite{2015SoPh..290.1399F} in which the solar wind during SC 23 is only analyzed.

The main point of the present study is to explore whether the proportions and properties of the solar wind originating from the same source region type are influenced by the solar cycle amplitude? Therefore, we concentrate on the comparison between the two solar cycles, SC 23 and SC 24. The proportions of three types of solar wind (CH, AR, and QS wind) in the same solar activity phases during SCs 23 and 24 are analyzed and compared. For simplicity, the solar minimum between SCs 23 and 24 (2007-2009) will be referred to as "the minimum phase of SC 23", and the solar minimum between SCs 24 and 25 (2018-2020) will be called "the minimum phase of SC 24".

The proportions of the same type of near-Earth solar wind during the two solar cycles are not the same. The proportions of AR and CH wind during SC 23 solar maximum and declining phases are higher than those during SC 24 maximum and declining phases. Whereas, the proportion of CH wind during SC 23 minimum is lower than that during SC 24 minimum. The proportions of CH wind during SC 23 (24) maximum, declining, and minimum phases are 30.91\% (26.23\%), 45.32\% (39.50\%), and 42.35\% (45.84\%), respectively. The proportions of AR wind during SC 23 (24) three phases are 46.59\% (42.41\%), 27.13\% (21.26\%), and 9.02\% (8.25\%), respectively. The proportions of QS wind during SC 23 maximum and declining (minimum) phases are lower (higher) than those during SC 24 maximum and declining (minimum) phases. The proportions of QS wind during SC 23 (24) solar maximum, declining, and minimum phases are 5.91\% (8.02\%), 12.04\% (17.34\%), and 44.57\% (40.81\%), respectively.

\begin{figure}[!h]
\centering
\includegraphics[width=0.7\textwidth]{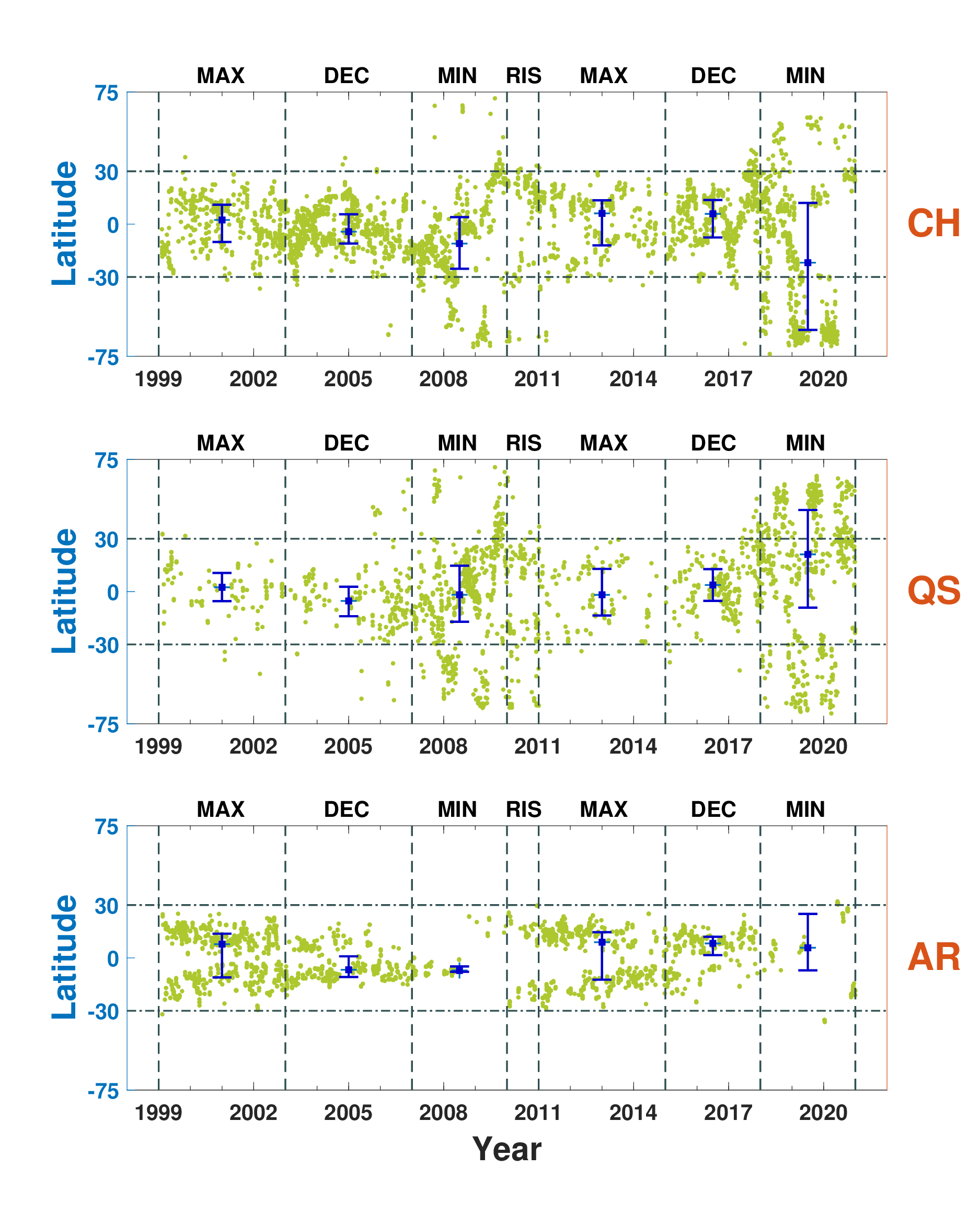}
\caption{The footpoint latitude distributions of CH, QS, and AR wind in each year during 1999-2020. The CH, QS, and AR wind are presented from the top to the bottom panels. The bars denote the median, upper, and lower quartiles of the footpoint latitudes during the three solar activity phases of SCs 23 and 24. The bars denote the median, upper, and lower quartiles of the footpoint latitudes during the maximum (1999-2002 and 2011-2014), declining (2003-2006 and 2015-2017), and minimum (2007-2009 and 2018-2020) phases of SCs 23 and 24.}
\label{footpoint_L}
\end{figure}

The present statistical results demonstrate that the source of the near-Earth solar wind is influenced by the solar cycle amplitude. The solar activity of SC 23 is significantly stronger than that of SC 24. The average yearly sunspot numbers of SC 23 maximum (161.1) and declining (58.8) phases are significantly higher than those of SC 24 maximum (93.2) and declining (39.1) phases. Previous studies demonstrated that the solar minimum of SC 23 is the weakest in nearly a century \citep{2009JGRA..114.9105G,2017ApJ...837..165R}. The average SSNs during solar minimum of 23 (2007-2009) is 7.2. The SSNs during the solar minimum of SC 24 (2018-2020) are also lower and similar to the previous solar minimum. The proportion differences in the three types of solar wind between SCs 23 and 24 can be explained reasonably by the solar cycle amplitude. The SSNs of maximum and declining phases during SC 23 are significantly higher than those of SC 24 (see top panel in Figure \ref{year}). The higher the SSNs, the more and stronger the active regions. Therefore, the proportions of AR wind during SC 23 maximum and declining phases are higher than those of SC 24. Generally, the near-Earth CH wind mainly comes from mid and low latitude coronal holes during solar maximum and decline phases (see top panel in Figure \ref{footpoint_L}). In addition, the mid and low latitude coronal holes are closely related to the active regions \citep{2009SSRv..144..383W}. As a result, strong solar activity not only leads to an increase in the number of active regions but also leads to an increase in the mid and low latitude coronal holes. This is consistent with the previous study that the area of coronal holes during the solar maximum of SC 23 is larger than that during the same phase of SC 24 \citep{2017SoPh..292...18L}. Hence the proportions of CH wind during SC 23 maximum and declining phases are higher than those of the same phases during SC 24. During SC 23 maximum and declining phases, there are more active regions and coronal holes compared with the SC 24 same phases. Therefore, there is less QS wind during SC 23 maximum and declining phases compared with the same phases during SC 24.

It is worth noting that the proportion of QS (CH) wind during the SC 23 minimum is higher (lower) than that during the SC 24 minimum. The proportions of QS (CH) wind during the minimums of SCs 23 and 24 are 44.57\% (42.35\%) and 40.81\% (45.84\%). To investigate the sources of the near-Earth solar wind in more detail, we show the footpoint latitude distributions for CH, AR, and QS wind in Figure \ref{footpoint_L}. The CH, QS, and AR wind are presented from the top to the bottom panels. The bars denote the median, upper, and lower quartiles of the footpoint latitudes during the maximum (1999-2002 and 2011-2014), declining (2003-2006 and 2015-2017), and minimum (2007-2009 and 2018-2020) phases of SCs 23 and 24. Generally, the footpoints of the solar wind are mainly located at the mid and low latitudes, ranging from the south to north 30 degrees.

The solar wind footpoint distribution characteristics during SCs 23 and 24 are also not the same. There are three characteristics of the footpoint distributions. First, the footpoint distribution of AR wind (bottom panel in Figure \ref{footpoint_L}) also presents the butterfly diagram, which is the same as that of sunspots. Second, the footpoint distribution ranges during SC 23 are smaller than those during SC 24 for all three types of solar wind. The median footpoint latitude for CH wind during SC 23 (24) maximum, declining, and minimum phases are 2.38 (6.10), -4.30 (5.83), and -11.05 (-21.88), respectively. The differences between the upper and lower quartiles of CH wind footpoint latitudes during SC 23 (24) maximum, declining, and minimum phases are 21.13 (25.61), 16.61 (21.26), and 29.30 (72.11), respectively. The AR and QS winds present the same characteristics as CH wind, the median and distribution ranges are shown in Table \ref{tab:table2}. Third, the footpoint distributions of SCs 23 and 24 solar minimums are significantly different. The near-Earth solar wind mainly comes from QS and CH regions during the two solar minimums (see bottom panel in Figure \ref{year}). The footpoint distributions of CH and QS winds during SC 23 minimum are generally symmetrical. In contrast, the distribution of the footpoints for CH (QS) wind is concentrated to the south (north) hemisphere during SC 24 minimum (see top and middle panels in Figure \ref{footpoint_L}). The median of the CH (QS) footpoint during SC 24 minimum is more than 20 degrees south (north), with the upper and lower quartiles of 12.07 and -60.04 (46.32 and -9.14).

\begin{table}[htbp]
  \centering
  \caption{The footpoint distributions for the three types of solar wind during solar maximum, declining, and minimum phases.}
    \begin{tabular}{ccccccc}
    \midrule
    \midrule
          &       &       & Median & Lower quartile & Upper quartile & Difference \\
    \multirow{6}[0]{*}{CH winds} & \multirow{3}[0]{*}{SC 23} & MAX   & 2.38  & -10.12 & 11.01 & 21.13 \\
          &       & DEC   & -4.30  & -11.00   & 5.61  & 16.61 \\
          &       & MIN   & -11.05 & -25.34 & 3.96  & 29.30 \\
          \cdashline{2-7}
          & \multirow{3}[0]{*}{SC 24} & MAX   & 6.10   & -12.06 & 13.55 & 25.61 \\
          &       & DEC   & 5.83  & -7.58 & 13.68 & 21.26 \\
          &       & MIN   & -21.88 & -60.04 & 12.07 & 72.11 \\
          \midrule
    \multirow{6}[0]{*}{AR winds} & \multirow{3}[0]{*}{SC 23} & MAX   & 7.79  & -11.07 & 13.74 & 24.81 \\
          &       & DEC   & -6.72 & -10.83 & 0.90   & 11.73 \\
          &       & MIN   & -7.14 & -7.86 & -4.86 & 3.00 \\
          \cdashline{2-7}
          & \multirow{3}[0]{*}{SC 24} & MAX   & 8.91  & -12.37 & 14.61 & 26.98 \\
          &       & DEC   & 8.23  & 1.58 & 12.00 & 10.42 \\
          &       & MIN   & 5.76  & -7.04 & 24.99 & 32.03 \\
          \midrule
    \multirow{6}[0]{*}{QS winds} & \multirow{3}[0]{*}{SC 23} & MAX   & 2.44  & -5.43 & 10.60  & 16.03 \\
          &       & DEC   & -5.30  & -14.00   & 2.71  & 16.71 \\
          &       & MIN   & -1.83 & -17.16 & 14.68 & 31.84 \\
          \cdashline{2-7}
          & \multirow{3}[0]{*}{SC 24} & MAX   & -1.85 & -13.57 & 12.87 & 26.44 \\
          &       & DEC   & 3.74  & -5.26 & 12.76 & 18.02 \\
          &       & MIN   & 21.12 & -9.14 & 46.32 & 55.46 \\
          \midrule
    \end{tabular}%
  \label{tab:table2}%
\end{table}%

The proportion and footpoint distribution differences of CH wind between SCs 23 and 24 solar minimums should be related to the larger structures of the corona. The present statistical results demonstrate that the proportion is lower and the footpoint distribution range is smaller for the CH wind during SC 23 minimum compared with those during SC 24 minimum. The previous study demonstrated that the polar coronal hole area during SC 23 minimum is 15\% smaller than that of the previous solar minimums \citep{2009SoPh..257...99K}. During SC 23 minimum, the solar wind originating from polar coronal holes is not easily transported to the Earth which is located at the ecliptic plane as the polar coronal holes shrink to the higher latitude. Therefore, the proportion of CH wind during the SC 23 minimum is lower than that during the SC 24 minimum. In addition, the CH wind is mainly coming from low latitudes during SC 23 minimum. In contrast, more CH winds originate from polar coronal holes during SC 24 minimum. Hence a lot of CH winds come from high latitudes and the proportion of CH wind during SC 24 minimum is higher than that during SC 23 minimum. Quanititively, during SC 23 (24) minimum, 21.99\% (55.03\%) of CH wind comes from regions with latitudes higher than 30 degrees (see the top panel in Figure \ref{footpoint_L}). The proportion differences of QS wind during SCs 23 and 24 minimums can also be explained by the larger structures of the corona. The above statistical results show that the proportion of QS wind is higher during SC 23 minimum compared with that during SC 24 minimum. Because the polar coronal holes shrink to the higher latitudes during SC 23 minimum. Therefore, the proportion of CH wind is lower and the proportion of QS wind is higher during SC 23 minimum.

\subsection{The comparison of the footpoint and in-situ magnetic field strengths, and speeds of the near-Earth solar wind between solar cycles 23 and 24}
\begin{figure}[!h]
\centering
\includegraphics[width=1.0\textwidth]{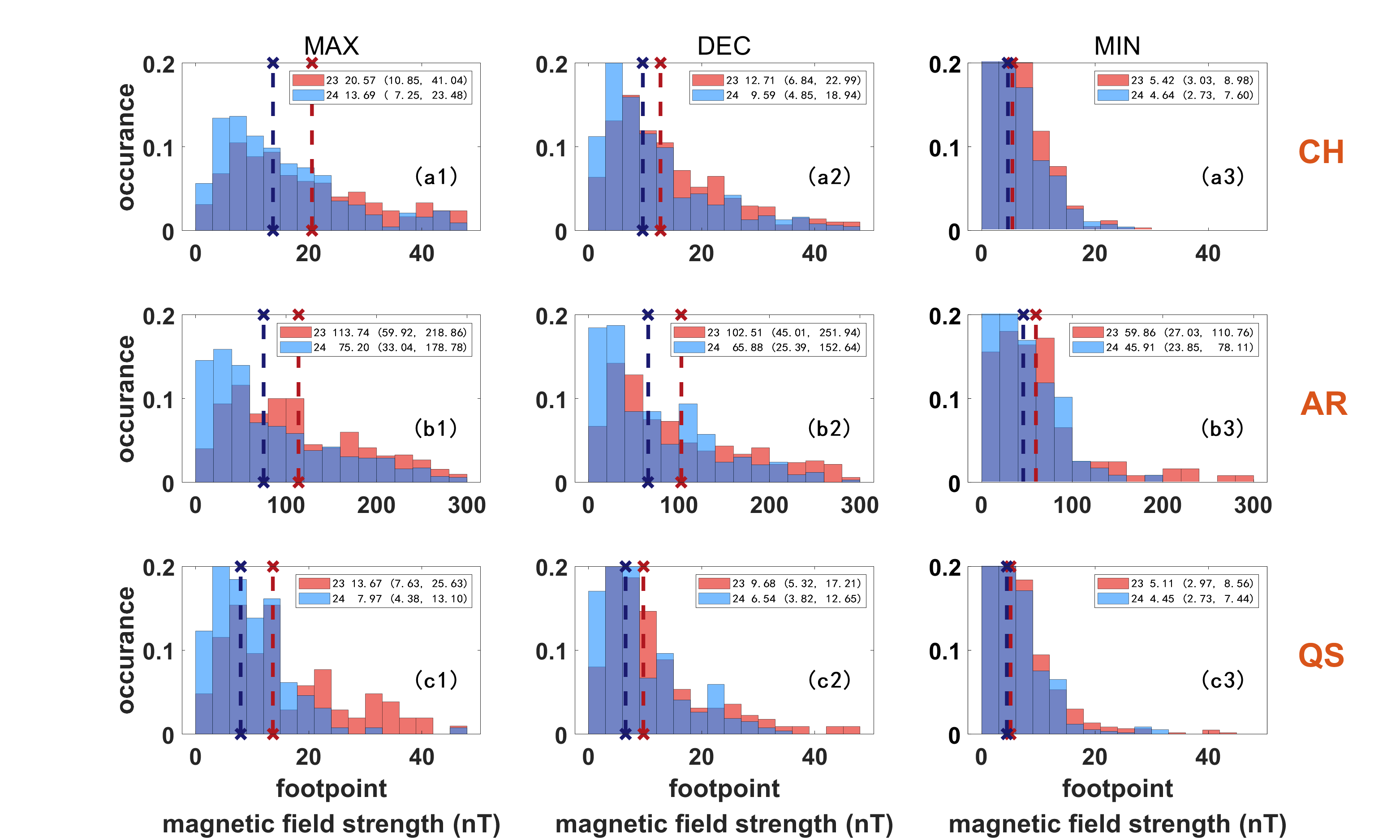}
\caption{The footpoint magnetic field strengths of three types of solar wind during SCs 23 (red) and 24 (blue). The dotted vertical lines denote the median footpoint magnetic field strengths. The magnetic field strength ranges (x-axis in the panels) for AR wind are larger as the magnetic field strengths of AR wind footpoints are much higher than those of QS and CH wind footpoints. The footpoint magnetic field strengths for all three types of solar wind are all higher during SC 23 maximum and declining phases than those during SC 24 maximum and declining phases.}
\label{acc_footpoint}
\end{figure}

The footpoint magnetic field strengths of the three types of solar wind during SCs 23 and 24 are presented in Figure \ref{acc_footpoint}. The statistical results are shown in solar maximum (left column), declining (middle column), and minimum (right column) phases, as the properties of solar wind change significantly with solar activity phases. The red and blue histograms represent SC 23 and SC 24. To make the introduction clearer, the statistical results are introduced in the following sequence. First, the properties of the three types of solar wind are simply compared. Second, the variations of the solar wind parameters with the solar activity phases are presented. Finally, the differences in the solar wind properties between SCs 23 and 24 are highlighted, as the present study concentrates on the influence of the solar cycle amplitude on the sources and properties of the solar wind.

The footpoint magnetic field strength of AR wind is much higher than that of QS and CH wind, and the footpoint magnetic field strength of CH wind is slightly higher than that of QS wind. The footpoint magnetic field strengths of the three types of solar wind all decrease from solar maximum to minimum.

It is worth noting that the footpoint magnetic field strengths of all three types of solar wind are all higher during SC 23 maximum and declining phases than those during SC 24 maximum and declining phases. In contrast, the footpoint magnetic field strengths of the three types of solar wind are close during the two solar minimums. During SC 23 (24) maximum, declining, and minimum phases, the median footpoint magnetic field strengths of CH wind are 20.57 (13.69), 12.71 (9.59), and 5.42 (4.64) nT, respectively. The median footpoint magnetic field strengths of QS wind are 13.67 (7.97), 9.68 (6.54), and 5.11 (4.45) nT during SC 23 (24) three solar activity phases. The median footpoint magnetic field strengths of AR wind are 113.74 (75.20), 102.51 (65.88), and 59.86 (45.91) nT during SC 23 (24) three activity phases.

Statistical results indicate that during the same phase in different solar cycles, the footpoint magnetic field strength of the solar wind is modulated by the SSNs. For the maximum and declining phases of SC 23, the SSNs are significantly higher than those in SC 24. Consequently, the footpoint magnetic field strengths of the solar wind in the corresponding periods of SC 23 are much higher than those in SC 24. During the minimum phases of the two solar cycles, the SSNs are similar and the footpoint magnetic field strengths of the same type solar wind are also comparable.

\begin{figure}[!h]
\centering
\includegraphics[width=1.0\textwidth]{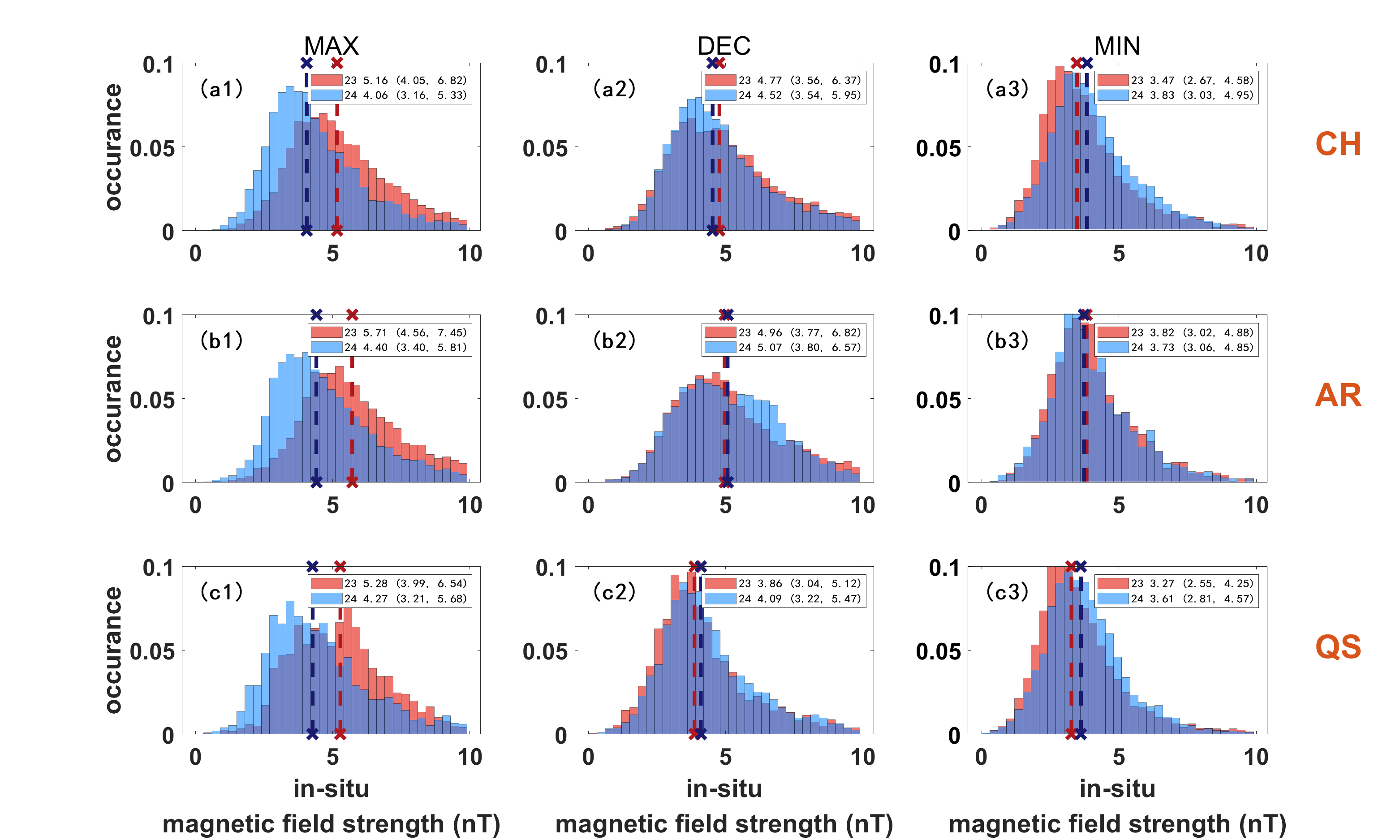}
\caption{The in-situ magnetic field strengths of three types of solar wind during SCs 23 (red) and 24 (blue). The dotted vertical lines denote the median in-situ magnetic field strength. Generally, the in-situ magnetic field strengths of AR and CH wind during SC 23 maximum and decline phases are higher than those of SC 24. The in-situ magnetic field strengths of CH and QS wind during SC 23 minimum are lower than those of SC 24.}
\label{acc_intu}
\end{figure}

The in-situ magnetic field strengths of the three types of solar wind during SCs 23 (red) and 24 (blue) are presented in Figure \ref{acc_intu}. There are only minor differences in the in-situ magnetic field strengths of the three types of solar wind. This is not the same as the footpoint magnetic field strengths of the three types of solar wind. The in-situ magnetic field strengths of all three types of solar wind also decrease from solar maximum to minimum. During the maximum and declining phases of SC 23, the in-situ magnetic field strengths of AR and CH wind are both higher than those in SC 24. This is clear for the solar maximum (see left in Figure \ref{acc_intu}). While during the solar minimum, it is the opposite, the in-situ magnetic field strengths of the CH and QS wind in SC 23 are both slightly lower than those in SC 24. The top panel of Figure \ref{acc_intu} gives the median in-situ magnetic field strengths of the CH wind during the maximum, declining, and minimum phases of SC 23 (24), which are 5.16 (4.06), 4.77 (4.52), and 3.47 (3.83) nT, respectively. The median in-situ magnetic field strengths of AR (middle panel) and QS (bottom panel) wind also show a clear dependence on the strength of solar activity. The median in-situ magnetic field strengths of the AR and QS wind during the maximum, declining, and minimum phases of the SC 23 (24) are as follows: for the AR wind, they are 5.71 (4.40), 4.96 (5.07), and 3.82 (3.73) nT, respectively; for the QS wind, they are 5.28 (4.27), 3.86 (4.09), and 3.27 (3.61) nT, respectively.

Our statistical results show that the in-situ magnetic field strengths of the three types of solar wind are very similar. In contrast, the footpoint magnetic field strengths of the AR wind are significantly higher than those of the CH and QS wind, approximately by an order of magnitude. This means that during the propagation from the Sun to the interplanetary, the AR wind undergoes a more pronounced super-radial expansion process and the degree of super-radial expansion is significantly higher than that of the CH and QS wind. In contrast, the degree of super-radial expansions of the CH and QS wind are similar. This is consistent with the notion of \cite{2009ApJ...691..760W} who suggested that the expansion factor of the AR wind is significantly higher than that of the CH wind.

Our present statistical results demonstrate that the in-situ magnetic field strengths of the three types of solar wind are all modulated by the solar cycle amplitude. To ensure statistical significance, we focus on all three solar phases for the CH wind, as there are sufficient CH wind samples in each solar phase. However, for the AR wind, we mainly focus on the solar maximum and declining phases, since there are very few AR wind during the solar minimum phase. Similarly, for the QS wind, we only focus on the solar minimum phase.

During the solar maximum and declining phases, the in-situ magnetic field strengths of CH and AR wind during SC 23 are higher than those during SC 24. Comparing the sunspot numbers between the two solar cycles, it is evident that the SC 23 has a significantly higher number of sunspots than SC 24 during the maximum and declining phases, resulting in a higher in-situ magnetic field strengths of the solar wind during those two phases in SC 23 than those in SC 24. In contrast, during the solar minimum phase, the in-situ magnetic field strengths of the CH and QS wind in SC 24 are higher than those in SC 23, although the SSNs of the two solar minimums are nearly the same. This indicates that the solar activities during the minimum phase of SC 23 are weaker than those of SC 24. This result also coincides with previous studies showing that the minimum phase of SC 23 is the weakest minimum in nearly a century. The present study indicates that the properties of solar wind may be the more sensitive index for representing solar activity. This is consistent with the results of \cite{2008GeoRL..3518103M}, in which the authors found that the speed, density, temperature, and dynamic pressure for high latitude solar wind during SC 23 minimum are all lower compared with those during SC 22 minimum.

\begin{figure}[!h]
\centering
\includegraphics[width=1.0\textwidth]{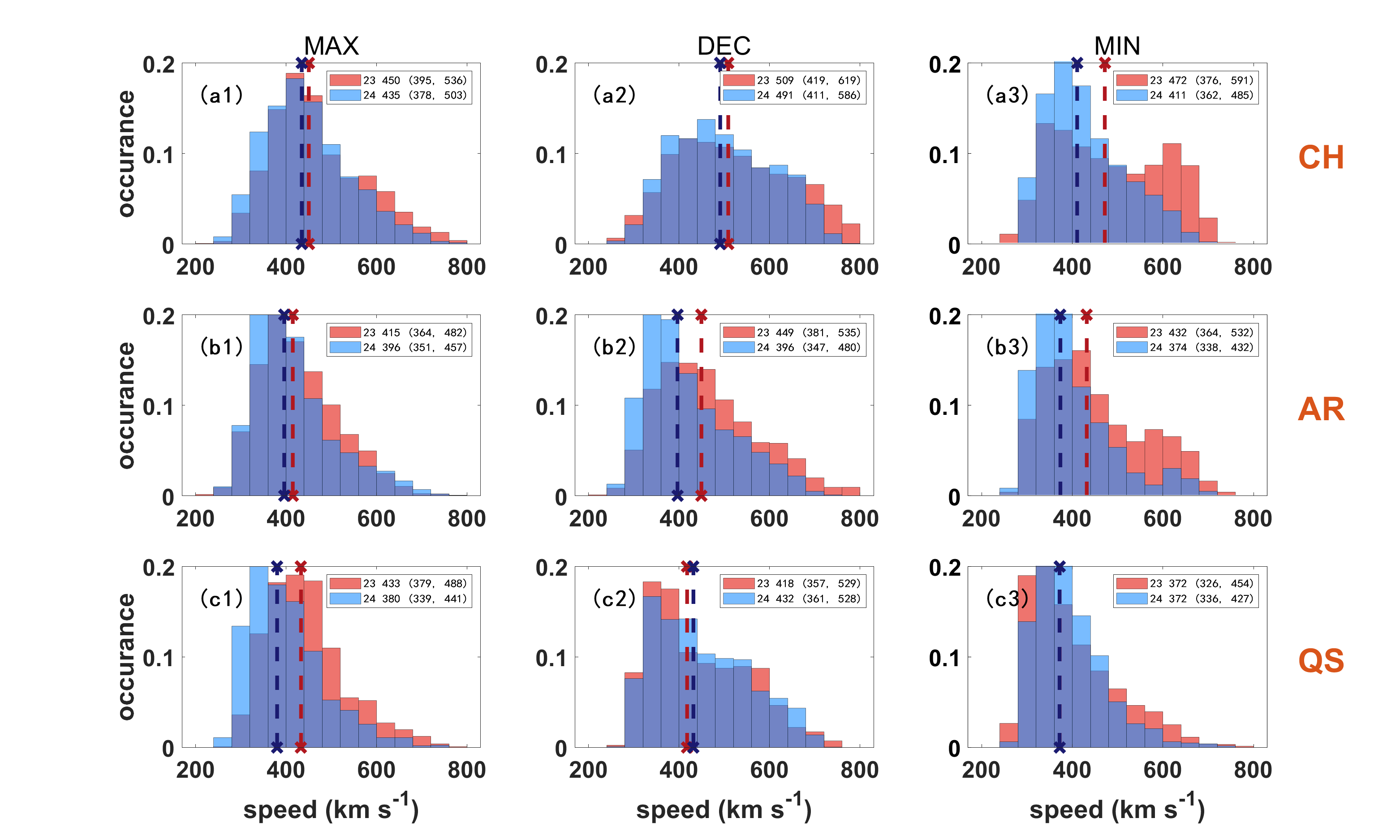}
\caption{The speeds of three types of solar wind during SCs 23 (red) and 24 (blue). The dotted vertical lines denote the median speeds.}
\label{acc_speed}
\end{figure}

The speeds of three types of solar wind during SCs 23 (red) and 24 (blue) are shown in Figure \ref{acc_speed}. The CH wind is faster than AR and QS wind. The speeds of the three types of solar wind are all influenced by solar activity phases, and they are all fastest during the solar declining phase. The speeds of all three types of solar wind during the maximum and declining phases of SC 23 are higher than those in SC 24, and the speeds of CH (QS) wind during SC 23 minimum are faster than (the same as) those during SC 24 minimum. The median speeds of CH wind during the solar maximum, declining, and minimum phases of the SC 23 (24) are 450 (435) \velunit, 509 (491) \velunit, and 472 (411) \velunit, respectively. The median speeds of AR wind during the solar maximum, declining, and minimum phases of the SC 23 (24) are 415 (396) \velunit, 449 (396) \velunit, and 432 (374) \velunit, respectively. The median speeds of QS wind during the three solar activity phases of the SC 23 (24) are 433 (380) \velunit, 418 (432) \velunit, and 372 (372) \velunit, respectively.

The present study demonstrates that the stronger the solar cycle amplitude, the higher the speeds of the solar wind originating from the same type of source region. During the maximum and declining phases of SC 23, the speeds of all three types of solar wind are higher than those during the same phases of SC 24. This indicates that the solar cycle amplitude can influence the speed of the solar wind, considering that the sunspot numbers are significantly higher during the maximum and declining phases of SC 23 than those in SC 24.

\begin{figure}[!h]
\centering
\includegraphics[width=1.0\textwidth]{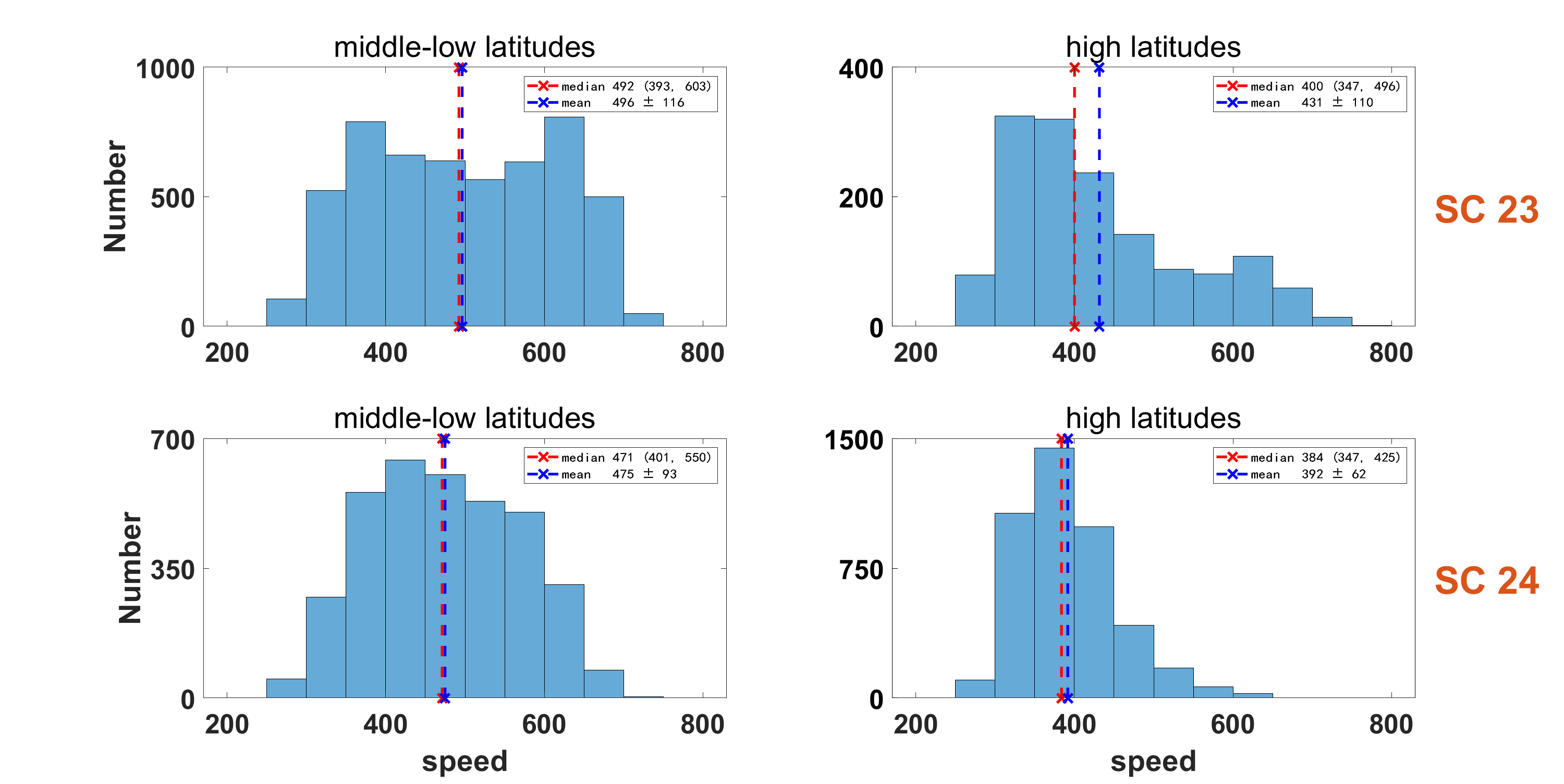}
\caption{The speed distributions of CH wind from middle-low latitudes (left column) and from high latitudes (right column) during the SC 23 (top row) and 24 (bottom row) minimum phases. The dotted vertical lines denote the median (red) and mean (blue) speeds. The speeds of CH wind from middle-low latitudes are higher than those from high latitudes.}
\label{acc_minimum}
\end{figure}

During the minimum phase, the speed of CH wind in SC 23 is significantly higher than that in SC 24, which may be related to the different latitudes of source regions during the two solar cycles. We further analyzed the speed distributions of CH wind from middle-low latitudes (below 30 degrees) and high latitudes (above 30 degress) during the SCs 23 and 24 minimum phases (see Figure \ref{acc_minimum}). The speed of CH wind from middle-low latitudes is notably higher than that of CH wind from high latitudes. During the SC 23 (24) minimum phases, the median speed of CH wind from middle-low latitudes is 492 (471) \velunit, while the median speed of CH wind from high latitudes is 400 (384) \velunit. This suggests that the solar wind from higher latitudes is associated with lower speed. The CH wind from high latitudes mainly originates from the boundaries of polar coronal holes, where the expansion factor is relatively large, resulting in slower solar wind speeds \citep{1990ApJ...355..726W}. Additionally, part of solar wind from high latitudes should be generated by the S-web model and be associated with slow speed. The above reasons may lead to the slower speeds of the CH wind from high latitudes. During the minimum phase of SC 23, the CH winds mainly (78.01\%) come from middle-low latitudes. In contrast, nearly half of the CH winds originate from higher latitudes during the minimum phase of SC 24. Additionally, during the minimum phase of SC 24, the polar coronal holes had larger areas with more open magnetic field lines. Therefore, during the solar minimum phase of SC 24, the solar wind from high latitudes is more likely generated by the S-web model and is associated with lower speed. This can reasonably explain the result that the speed of CH wind during the minimum phase of SC 23 is higher than that of SC 24.

During the SC 23 minimum phase, CH wind primarily originated from middle-low latitudes coronal holes, which may contribute to the double-peak distribution of CH wind speeds during the SC 23 minimum. The left peak with a lower speed should originate from coronal hole boundaries. In contrast, the right peak with a higher speed should come from the center regions of coronal holes. It should be noted that the double-peak distribution of CH wind speeds during the SC 23 minimum may also be due to inaccurate classification of solar wind. Due to errors introduced by the tracking method and the inherent inaccuracies of the PFSS model, each type of solar wind may include other types of solar wind. Therefore, the double-peak distribution observed in CH wind speeds during the SC 23 minimum could also be attributed to classification inaccuracies.

\subsection{The comparison of the charge states and helium abundance in the near-Earth solar wind between solar cycles 23 and 24}
\begin{figure}[!h]
\centering
\includegraphics[width=1.0\textwidth]{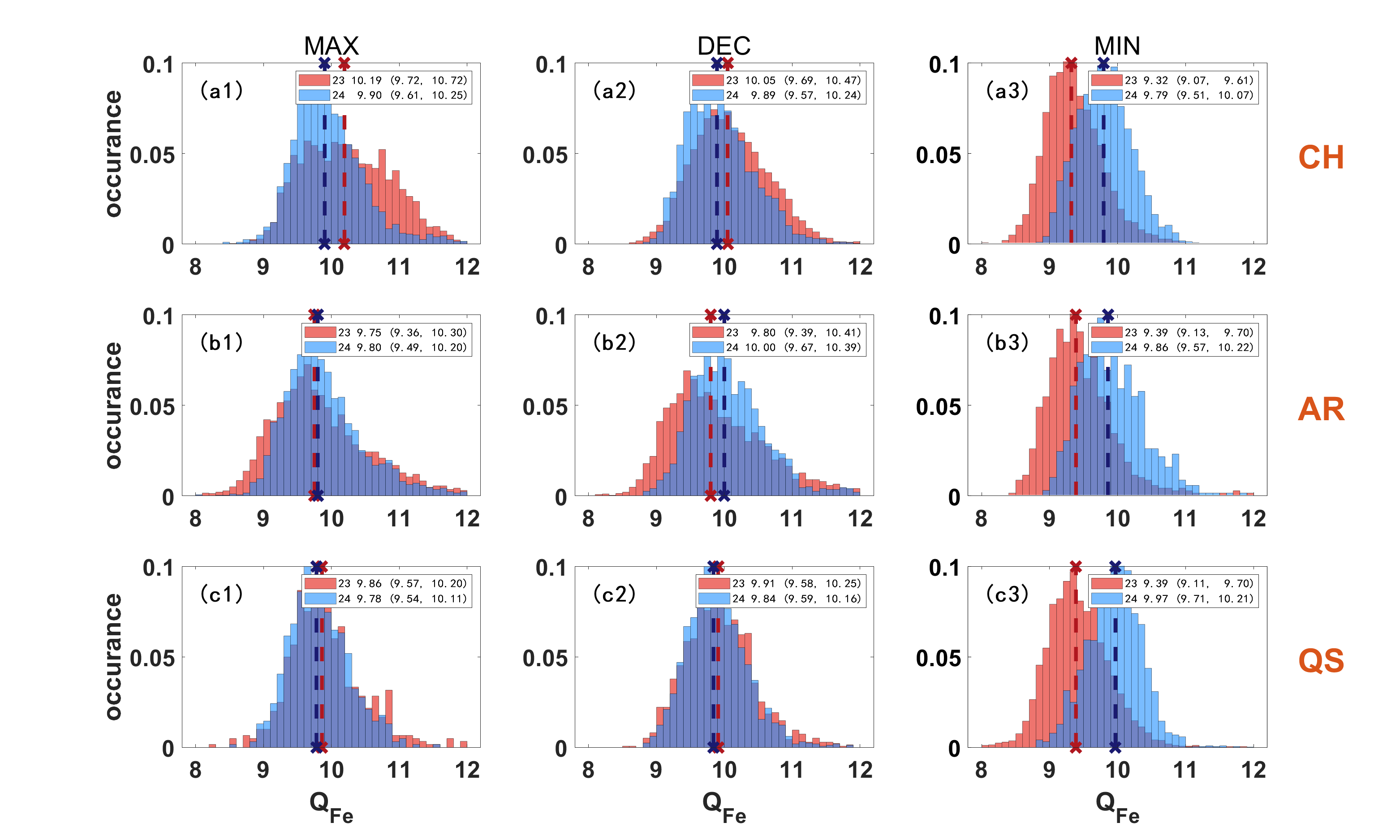}
\caption{The $Q_{Fe}$ in three types of solar wind during SCs 23 (red) and 24 (blue). The dotted vertical lines denote the median $Q_{Fe}$. The most noticeable characteristic is that the $Q_{Fe}$ in three types of solar wind are all extremely lower during SC 23 solar minimum.}
\label{acc_QFe}
\end{figure}

The average charge states of iron ($Q_{Fe}$) in the three types of solar wind during SCs 23 (red) and 24 (blue) are presented in  Figure \ref{acc_QFe}. In short, the differences in $Q_{Fe}$ among the three types of solar wind are not very significant. The most noticeable characteristic is that the $Q_{Fe}$ in the three types of solar wind are all extremely lower during the SC 23 solar minimum. During the maximum, declining, and minimum phases of SC 23 (24), the median $Q_{Fe}$ in CH wind are 10.19 (9.90), 10.05 (9.89), and 9.32 (9.79), respectively; the median $Q_{Fe}$ in AR wind are 9.75 (9.80), 9.80 (10.00), and 9.39 (9.86), respectively; the median $Q_{Fe}$ in QS wind are 9.86 (9.78), 9.91 (9.84), and 9.39 (9.97), respectively. During the SC 24, the median $Q_{Fe}$ in the three types of solar wind show minor differences across the three solar phases, ranging from around 9.8 to 10.0. However, there is a notable decrease in the median $Q_{Fe}$ in all three types of solar wind during the minimum phase of SC 23, dropping to around 9.3-9.4. In contrast, the $Q_{Fe}$ during the SC 24 minimum is nearly the same as that during the SC 24 maximum and declining phases (see left panel in Figure \ref{acc_QFe}).

\begin{figure}[!h]
\centering
\includegraphics[width=1.0\textwidth]{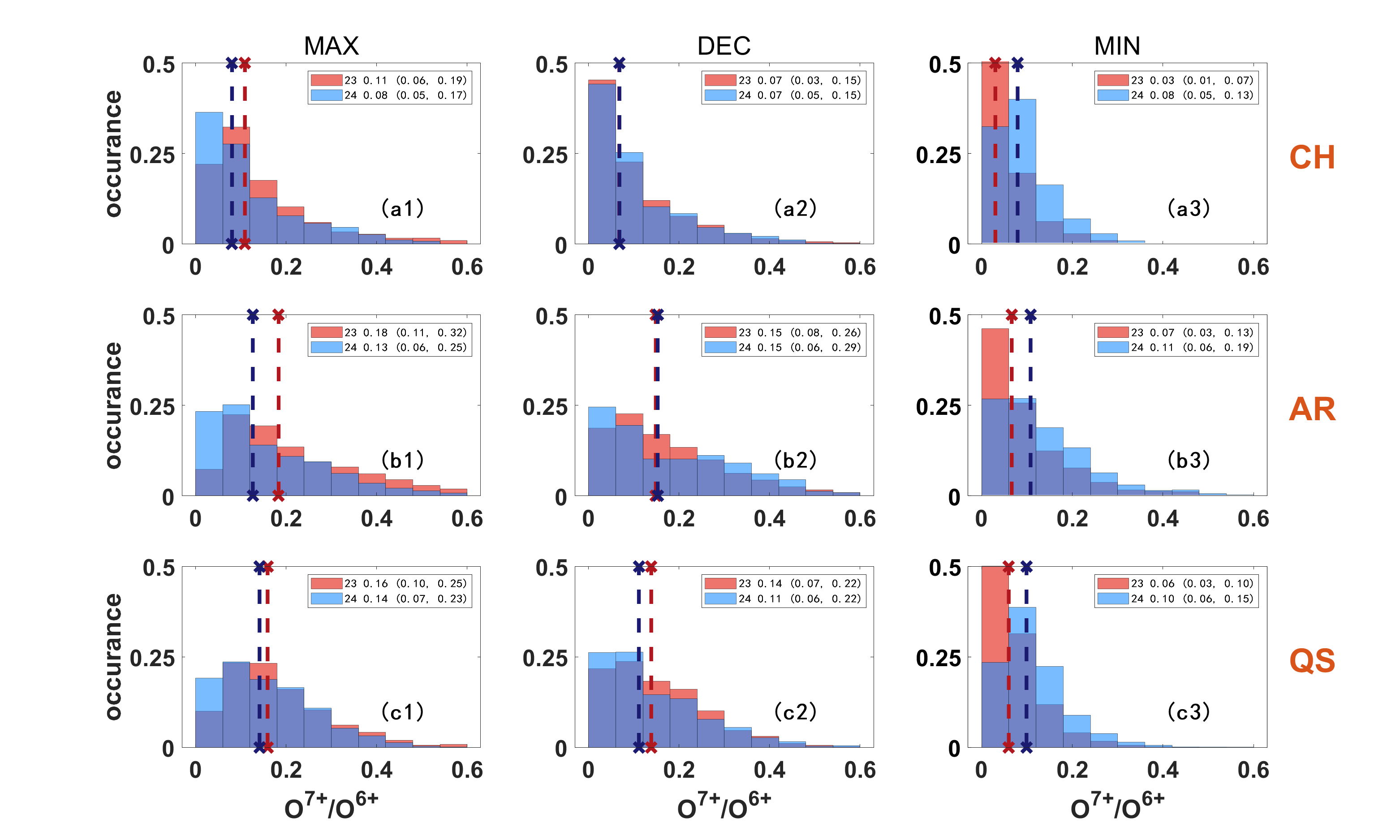}
\caption{The $O^{7+}/O^{6+}$ in three types of solar wind during SCs 23 (red) and 24 (blue). The dotted vertical lines denote the median $O^{7+}/O^{6+}$. The most noticeable characteristic is that the $O^{7+}/O^{6+}$ in three types of solar wind are all extremely lower during SC 23 solar minimum.}
\label{acc_O76}
\end{figure}

The density ratio of $O^{7+}$ and $O^{6+}$ ($O^{7+}/O^{6+}$) in the three types of solar wind during SCs 23 (red) and 24 (blue) are shown in Figure \ref{acc_O76}. Different from the $Q_{Fe}$, which is similar in the three types of solar wind, the $O^{7+}/O^{6+}$ is the highest in AR wind and lowest in CH wind, with QS wind lying in between. The $O^{7+}/O^{6+}$ in the three types of solar wind all decrease significantly from solar maximum to minimum during the SC 23. In contrast, during the SC 24, the $O^{7+}/O^{6+}$ in the three types of solar wind all decrease slightly from solar maximum to minimum. Comparing the two solar cycles, the $O^{7+}/O^{6+}$ in all three types of solar wind during the SC 23 solar maximum and declining phases are all higher than those during the same phases of SC 24. In contrast, the $O^{7+}/O^{6+}$ in the three types of solar wind during the SC 23 minimum are lower than those during the SC 24 minimum. During the maximum, declining, and minimum phases of SC 23 (24), the median $O^{7+}/O^{6+}$ in CH wind are 0.11 (0.08), 0.07 (0.07), and 0.03 (0.08), respectively; the median $O^{7+}/O^{6+}$ in AR wind are 0.18 (0.13), 0.15 (0.15), and 0.07 (0.11), respectively; the median $O^{7+}/O^{6+}$ in QS wind are 0.16 (0.14), 0.14 (0.11), and 0.06 (0.10), respectively.

The above results on charge states of $Q_{Fe}$ and $O^{7+}/O^{6+}$ indicate that the temperature distribution characteristics of the corona are not the same at different altitudes. Our statistical results show that there is a significant difference in $O^{7+}/O^{6+}$ in the three types of solar wind, the $O^{7+}/O^{6+}$ in AR wind is the highest, CH wind is the lowest, and QS wind lies in between. However, the $Q_{Fe}$ in the solar wind originating from different types of source regions is nearly the same. Previous simulations have shown that the freezing heights of the two elements are different, the charge states of oxygen freeze at about 0.5 R$_{\bigodot}$, and the charge states of iron freeze at about 3-5 R$_{\bigodot}$ \citep{1997SoPh..171..345K,1998ApJ...498..448E,2001ApJ...563.1055E}. Qualitatively, the charge states of solar wind represent the temperature when the freezing effect occurs. Therefore, our statistical results mean that there is a significant difference in the temperatures of the AR, QS, and CH regions in the low atmosphere. In contrast, the temperatures of the corona may be nearly the same at the altitude of 3-5 R$_{\bigodot}$. This is consistent with the observations taken by \cite{2010ApJ...708.1650H,2013SoPh..285....9H,2021ApJ...911L...4H} during solar eclipses. It is clear that the temperatures of AR, QS, and CH regions are quite different at low altitudes. However, the temperatures of above three types of regions are almost uniform at high altitudes (please see Figure~2, 4, and 5 in \cite{2021ApJ...911L...4H}). Additionally, our statistical results demonstrate that the minimum of SC 23 is extremely weak, as the charge states for all three types of solar wind during SC 23 minimum are significantly lower than those of SC 24 minimum.

\begin{figure}[!h]
\centering
\includegraphics[width=1.0\textwidth]{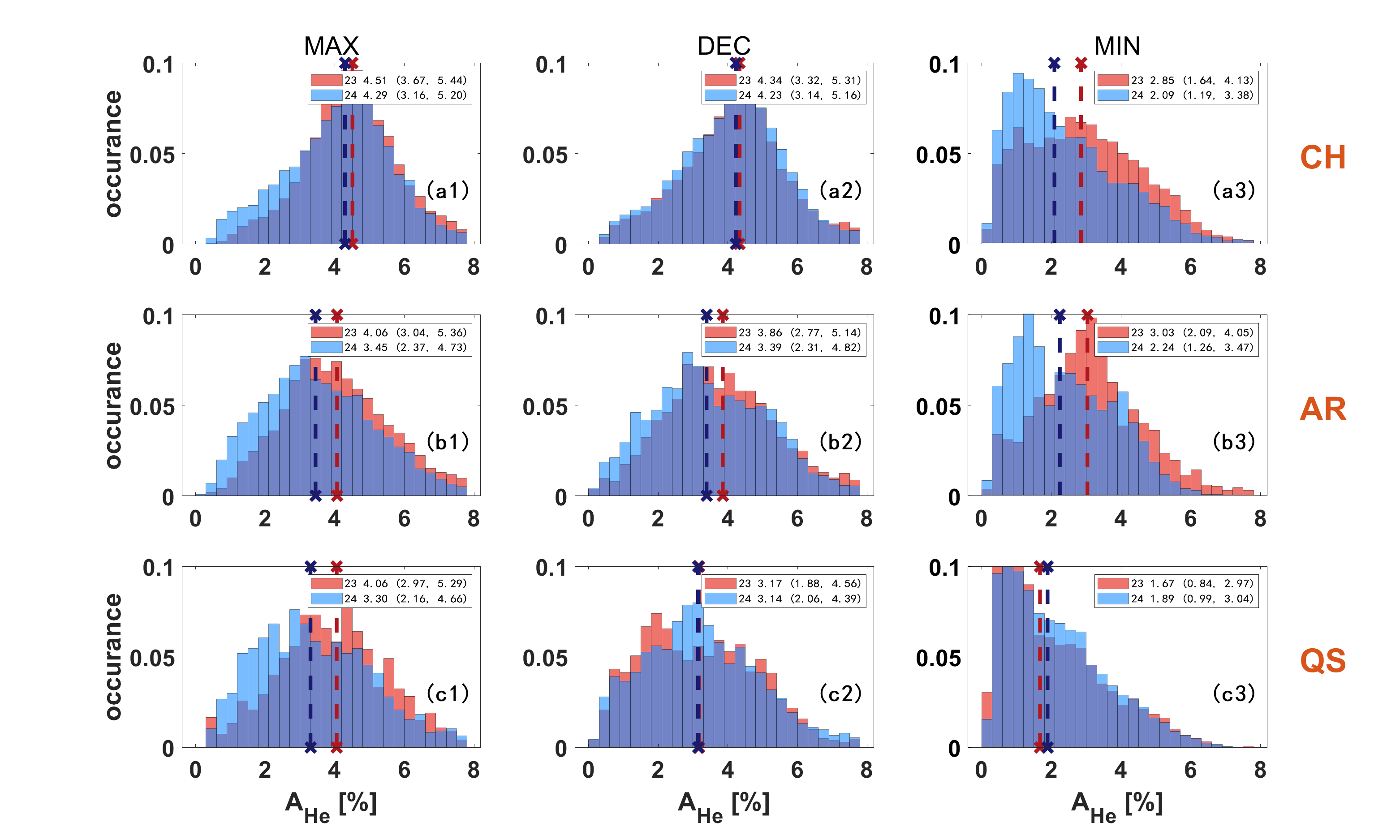}
\caption{The $A_{He}$ in three types of solar wind during SCs 23 (red) and 24 (blue). The dotted vertical lines denote the median $A_{He}$.}
\label{acc_AHe}
\end{figure}

The $A_{He}$ in three types of solar wind during SCs 23 (red) and 24 (blue) are shown in Figure \ref{acc_AHe}. Generally, the $A_{He}$ is highest in CH wind, and lowest in QS wind, with AR wind lying in between. The $A_{He}$ in the three types of solar wind all decrease from solar maximum to minimum. During solar maximum and declining phases, the $A_{He}$ is higher during SC 23 than that during SC 24 for all three types of solar wind. The $A_{He}$ of CH (QS) wind during SC 23 minimum is higher (lower) than that of SC 24 minimum. During SC 23 (24) maximum, declining, and minimum phases, the median $A_{He}$ in CH wind are 4.51 (4.29), 4.34 (4.23), and 2.85 (2.09), respectively; the median $A_{He}$ in AR wind are 4.06 (3.45), 3.86 (3.39), and 3.03 (2.24), respectively; and the median $A_{He}$ in QS wind are 4.06 (3.30), 3.17 (3.14), and 1.67 (1.89), respectively.

The statistical results of $A_{He}$ in the three types of solar wind indicate that the stronger solar activity is more likely to produce higher $A_{He}$. During solar maximum and declining phases, the SSNs in SC 23 are much higher than those in SC 24. Therefore, the $A_{He}$ in all three types of solar wind during SC 23 maximum and declining phases are higher than those during SC 24. However, the $A_{He}$ in the solar wind during the minimum phase is more complicated. The $A_{He}$ in QS wind during SC 23 minimum is lower than that during SC 24. This is consistent with the fact that the solar activity during the SC 23 minimum is lower than that during the SC 24 minimum.

The solar wind generated by the S-web model can also connected with the solar wind differences between the two solar minimums. According to the S-Web model, the slow wind comes from the continuous opening and closing of narrow open-field corridors with an intrinsic dynamic process \citep{2011ApJ...731..112A,2012SSRv..172..169A,2016SSRv..201...55A,2023NatAs...7..133C,2023ApJ...950...65B}. In this scenario, the plasma of the solar wind comes from the middle corona with heights of about 1.5 R$_{\bigodot}$ \citep{2016SSRv..201...55A,2017ApJ...840L..10H,2023NatAs...7..133C}. During solar minimum, the slow solar wind can be produced by the S-Web model, in which the dynamic rather than quasi-static. During the solar minimum phases of the two solar cycles, the areas of polar coronal holes in SC 23 are smaller than those in SC 24. Therefore, during the solar minimum phase of SC 24, the larger polar coronal hole area may result in a more extensive open magnetic field region with more open magnetic field lines. This could lead to the formation of more or wider magnetic corridors, thereby increasing the amount of slow solar wind. Additionally, the solar wind generated by the S-Web model should associated with lower $A_{He}$ as the helium abundance in the higher corona is lower than that of the lower atmosphere \citep{2009JGRA..114.4103S}. During the solar minimum phase of SC 24, a larger portion of the CH wind is more likely to be generated through the S-web model. Therefore, the speed and $A_{He}$ of CH wind during SC 24 minimum are both lower than those during the solar minimum phase of SC 23.

Previous studies have shown that the minimum of SC 23 is longer and its SSNs are lower compared with the previous minimums in nearly a century \citep{2009JGRA..114.9105G,2017ApJ...837..165R}. Hence, there are significant differences in coronal structures and the properties of solar wind in different latitudes between the minimum of the SC 23 and the previous minimum phases. The properties of the fast solar wind originating from large polar coronal holes measured by Ulysses during the minimums of SCs 22 and 23 are not the same. \cite{2008GeoRL..3522103S} found that the radial heliospheric magnetic field strength is lower during the minimum of SC 23 than that during the minimum of SC 22. The statistical results demonstrate that the fast solar wind originating from large polar coronal holes during the minimum of SC 23 is associated with lower speed, density, temperature, and dynamic pressure compared with the previous solar minimum \citep{2008GeoRL..3518103M,2008GeoRL..3519101I}. In addition, the parameters of the solar wind near the Earth during the minimum of SC 23 are also different from the previous minimums. The heliospheric magnetic field strength near the Earth is lower during the minimum of SC 23 than that of the previous minimums \citep{2008GeoRL..3520108O,2009SoPh..256..345L}. \cite{2015JGRA..12010250Z} compared the parameters of the solar wind detected near the Earth over the four solar minimums. They found that the magnetic field strength, density, and speed during the minimum of SC 23 are all lower compared with the previous solar minimums. The above statistical results show that the minimum of SC 23 is indeed very weak. Now we know that the solar activity of SC 24 is also extremely weaker compared to the previous solar cycles in the recent century. This means that the lowest solar cycle amplitude follows the weakest solar minimum in the recent century.

The present study indicates that the properties of the solar wind during solar minimum might be used to infer the amplitude of the following solar cycle. We find that the in-situ magnetic field strengths and charge states of the three types of solar wind during SC 24 minimum are generally higher than those during SC 23 minimum, although there is not a significant difference in sunspot numbers between the two minimums. Consistently, the amplitude of SC 25 (following the SC 24 minimum) is also stronger than that of SC 24 (following the SC 23 minimum). The property differences are more significant in the charge states between the two minimums. The charge states of the solar wind during SC 23 minimum are significantly lower than those during SC 24 minimum. The $Q_{Fe}$ for CH, AR, and QS solar wind is 9.32 (9.79), 9.39 (9.86), and 9.39 (9.97) during SC 23 (24) minimum. The $O^{7+}/O^{6+}$ for CH, AR, and QS solar wind is 0.03 (0.08), 0.07 (0.11), and 0.06 (0.10) during SC 23 (24) minimum. Our statistical results indicate that the charge states of the solar wind may be a more sensitive indicator of solar activity. Previous studies have shown that the response of the higher atmospheric layers to solar activity is more significant. For instance, in the lower atmosphere, the radiation from the photosphere differs by only 0.1\% between solar maximum and solar minimum \citep{2003ESASP.535..183F,2005AdSpR..35..376S}, while the radiation from the chromosphere varies by several tens of percentages \citep{1986JGR....91.8672H,2000JGR...10527195W,2013SSRv..176..237F}. The radiation from the outermost atmospheric layer, the solar corona, varies several times or more between solar maximum and solar minimum \citep{1994SoPh..152...53A,2005JGRA..110.8106R,2014Ap&SS.350..479R}. However, we cannot continuously and reliably measure various properties of the solar corona, such as the temperature and elemental abundance. In addition, the coronal structures are highly non-uniform, so it is challenging to provide an accurate description of how coronal properties evolve with solar activity. The solar wind carries information from the Sun, and can be used to infer the properties of the solar atmosphere. Therefore, the solar wind properties during the minimum phase may be used to infer the solar cycle amplitude. Certainly, more observations are needed for the above speculation.

\section{Summary and Conclusions} \label{sec:displaymath}
In the present study, we use the improved two-step mapping procedure to trace the near-Earth solar wind back to its source region during 1999-2020. The statistical results demonstrate that our improved method can link the near-Earth solar wind to its source region accurately. Subsequently, we classify the solar wind into three types: coronal hole, active region, and quiet Sun solar wind based on the characteristics of source regions. We concentrate on the comparison of sources and properties of the solar wind between SCs 23 and 24. The statistical results demonstrate that the solar cycle amplitude influences the sources and properties of the solar wind. The main results are as follows:
\begin{enumerate}

\item
The sources and footpoint latitude distributions of the three types of solar wind are all influenced by solar cycle amplitude. During the maximum and declining phases, the proportions of CH and AR wind are higher in SC 23 than those in SC 24. While during the solar minimum, the proportion of QS wind is higher in SC 24. In addition, the footpoint latitude distribution ranges during SC 23 are smaller than those during SC 24 for all three types of solar wind. It is worth noting that during the SC 24 minimum, the proportion of CH wind from high latitudes is much higher than that during the SC 23 minimum.

\item
The footpoint and in-situ magnetic field strengths of the three types of solar wind are also affected by solar cycle amplitude. During the solar maximum and declining phases, the footpoint and in-situ magnetic field strengths of the three types of solar wind are all higher during SC 23. During the solar minimum, the above parameters of the three types of solar wind are higher during SC 24 or similar during the two solar cycles. The in-situ magnetic field strengths of the three types of solar wind are similar, but the footpoint magnetic field strength of AR wind is much higher than that of CH and QS wind. During the solar maximum and declining phases, the speeds of the three types of solar wind are all higher during SC 23 than those during SC 24.

\item
Generally, the charge states ($Q_{Fe}$ and $O^{7+}/O^{6+}$) and $A_{He}$ in the three types of solar wind during the solar maximum and declining phases of SC 23 are all higher than those during the same solar phases of SC 24. However, during the solar minimum, the charge states during SC 23 have a sharp drop compared with other phases of SC 23, and they are also significantly lower than those during SC 24 minimum. The $O^{7+}/O^{6+}$ in AR wind is significantly higher than that of the CH wind, with the $O^{7+}/O^{6+}$ in QS wind lying in between. While the $Q_{Fe}$ in the three types of solar wind is nearly the same.

\end{enumerate}

Our statistical results demonstrate that the sources and properties of the solar wind are both influenced by the solar cycle amplitude. During the maximum and declining phases, the sunspot numbers of SC 23 are much higher than those of SC 24. Therefore, the speeds, footpoint and in-situ magnetic field strengths, $Q_{Fe}$, $O^{7+}/O^{6+}$, and $A_{He}$ of all three types of solar wind are generally higher during SC 23 than those during SC 24. In contrast, our statistical analyses demonstrate that the above solar parameters during SC 24 minimum are higher or equal to those during SC 23 minimum, although the sunspot numbers are nearly the same during the minimum phases of SCs 23 and 24.
The above results indicate that the charge states of the solar wind may be a more sensitive indicator for the intensity of solar activity.
In addition, our statistical results mean that there is a significant difference in the temperatures of the AR, QS, and CH regions at the lower atmosphere. In contrast, the temperatures of the above three types of regions are almost uniform at the altitude of 3-5 R$_{\bigodot}$.


\begin{acknowledgments}

The authors thank the anonymous referee for the helpful and constructive comments and suggestions. We thank the ACE SWICS instrument team and the ACE Science Center for providing the ACE
data. Analysis of Wind SWE observations is supported by NASA grant NNX09AU35G. SOHO is a project of international cooperation between ESA and NASA. SDO is a mission
of NASAs Living With a Star Program. The yearly SSNs are provided by the Solar Influence Data Center of the Royal Observatory of Belgium. This research is supported by the National Natural Science Foundation of China (42230203, 42074207).
\end{acknowledgments}

\bibliography{references1}{}

\begin{thebibliography}{}
\expandafter\ifx\csname natexlab\endcsname\relax\def\natexlab#1{#1}\fi
\providecommand{\url}[1]{\href{#1}{#1}}
\providecommand{\dodoi}[1]{doi:~\href{http://doi.org/#1}{\nolinkurl{#1}}}
\providecommand{\doeprint}[1]{\href{http://ascl.net/#1}{\nolinkurl{http://ascl.net/#1}}}
\providecommand{\doarXiv}[1]{\href{https://arxiv.org/abs/#1}{\nolinkurl{https://arxiv.org/abs/#1}}}

\bibitem[{{Abbo} {et~al.}(2016){Abbo}, {Ofman}, {Antiochos}, {Hansteen},
  {Harra}, {Ko}, {Lapenta}, {Li}, {Riley}, {Strachan}, {von Steiger}, \&
  {Wang}}]{2016SSRv..201...55A}
{Abbo}, L., {Ofman}, L., {Antiochos}, S.~K., {et~al.} 2016, \ssr, 201, 55,
  \dodoi{10.1007/s11214-016-0264-1}

\bibitem[{{Acu{\~n}a} {et~al.}(1995){Acu{\~n}a}, {Ogilvie}, {Baker}, {Curtis},
  {Fairfield}, \& {Mish}}]{1995SSRv...71....5A}
{Acu{\~n}a}, M.~H., {Ogilvie}, K.~W., {Baker}, D.~N., {et~al.} 1995, \ssr, 71,
  5, \dodoi{10.1007/BF00751323}

\bibitem[{{Aellig} {et~al.}(2001){Aellig}, {Lazarus}, \&
  {Steinberg}}]{2001GeoRL..28.2767A}
{Aellig}, M.~R., {Lazarus}, A.~J., \& {Steinberg}, J.~T. 2001, \grl, 28, 2767,
  \dodoi{10.1029/2000GL012771}

\bibitem[{{Alterman} \& {Kasper}(2019)}]{2019ApJ...879L...6A}
{Alterman}, B.~L., \& {Kasper}, J.~C. 2019, \apjl, 879, L6,
  \dodoi{10.3847/2041-8213/ab2391}

\bibitem[{{Altschuler} \& {Newkirk}(1969)}]{1969SoPh....9..131A}
{Altschuler}, M.~D., \& {Newkirk}, G. 1969, \solphys, 9, 131,
  \dodoi{10.1007/BF00145734}

\bibitem[{{Antiochos} {et~al.}(2012){Antiochos}, {Linker}, {Lionello},
  {Miki{\'c}}, {Titov}, \& {Zurbuchen}}]{2012SSRv..172..169A}
{Antiochos}, S.~K., {Linker}, J.~A., {Lionello}, R., {et~al.} 2012, \ssr, 172,
  169, \dodoi{10.1007/s11214-011-9795-7}

\bibitem[{{Antiochos} {et~al.}(2011){Antiochos}, {Miki{\'c}}, {Titov},
  {Lionello}, \& {Linker}}]{2011ApJ...731..112A}
{Antiochos}, S.~K., {Miki{\'c}}, Z., {Titov}, V.~S., {Lionello}, R., \&
  {Linker}, J.~A. 2011, \apj, 731, 112, \dodoi{10.1088/0004-637X/731/2/112}

\bibitem[{{Aschwanden}(1994)}]{1994SoPh..152...53A}
{Aschwanden}, M.~J. 1994, \solphys, 152, 53, \dodoi{10.1007/BF01473183}

\bibitem[{{Baker} {et~al.}(2023){Baker}, {D{\'e}moulin}, {Yardley},
  {Mihailescu}, {van Driel-Gesztelyi}, {D'Amicis}, {Long}, {To}, {Owen},
  {Horbury}, {Brooks}, {Perrone}, {French}, {James}, {Janvier}, {Matthews},
  {Stangalini}, {Valori}, {Smith}, {Cuadrado}, {Peter}, {Schuehle}, {Harra},
  {Barczynski}, {Berghmans}, {Zhukov}, {Rodriguez}, \&
  {Verbeeck}}]{2023ApJ...950...65B}
{Baker}, D., {D{\'e}moulin}, P., {Yardley}, S.~L., {et~al.} 2023, \apj, 950,
  65, \dodoi{10.3847/1538-4357/acc653}

\bibitem[{{Buergi} \& {Geiss}(1986)}]{1986SoPh..103..347B}
{Buergi}, A., \& {Geiss}, J. 1986, \solphys, 103, 347,
  \dodoi{10.1007/BF00147835}

\bibitem[{{Cane} \& {Richardson}(2003)}]{2003JGRA..108.1156C}
{Cane}, H.~V., \& {Richardson}, I.~G. 2003, Journal of Geophysical Research
  (Space Physics), 108, 1156, \dodoi{10.1029/2002JA009817}

\bibitem[{{Chi} {et~al.}(2016){Chi}, {Shen}, {Wang}, {Xu}, {Ye}, \&
  {Wang}}]{2016SoPh..291.2419C}
{Chi}, Y., {Shen}, C., {Wang}, Y., {et~al.} 2016, \solphys, 291, 2419,
  \dodoi{10.1007/s11207-016-0971-5}

\bibitem[{{Chitta} {et~al.}(2023){Chitta}, {Seaton}, {Downs}, {DeForest}, \&
  {Higginson}}]{2023NatAs...7..133C}
{Chitta}, L.~P., {Seaton}, D.~B., {Downs}, C., {DeForest}, C.~E., \&
  {Higginson}, A.~K. 2023, Nature Astronomy, 7, 133,
  \dodoi{10.1038/s41550-022-01834-5}

\bibitem[{{Cliver} \& {Ling}(2001)}]{2001ApJ...556..432C}
{Cliver}, E.~W., \& {Ling}, A.~G. 2001, \apj, 556, 432, \dodoi{10.1086/321570}

\bibitem[{{Cranmer}(2009)}]{2009LRSP....6....3C}
{Cranmer}, S.~R. 2009, Living Reviews in Solar Physics, 6, 3,
  \dodoi{10.12942/lrsp-2009-3}

\bibitem[{{Cranmer} {et~al.}(2007){Cranmer}, {van Ballegooijen}, \&
  {Edgar}}]{2007ApJS..171..520C}
{Cranmer}, S.~R., {van Ballegooijen}, A.~A., \& {Edgar}, R.~J. 2007, \apjs,
  171, 520, \dodoi{10.1086/518001}

\bibitem[{{Delaboudini{\`e}re} {et~al.}(1995){Delaboudini{\`e}re}, {Artzner},
  {Brunaud}, {Gabriel}, {Hochedez}, {Millier}, {Song}, {Au}, {Dere}, {Howard},
  {Kreplin}, {Michels}, {Moses}, {Defise}, {Jamar}, {Rochus}, {Chauvineau},
  {Marioge}, {Catura}, {Lemen}, {Shing}, {Stern}, {Gurman}, {Neupert},
  {Maucherat}, {Clette}, {Cugnon}, \& {Van Dessel}}]{1995SoPh..162..291D}
{Delaboudini{\`e}re}, J.~P., {Artzner}, G.~E., {Brunaud}, J., {et~al.} 1995,
  \solphys, 162, 291, \dodoi{10.1007/BF00733432}

\bibitem[{{Dicke}(1979)}]{1979Natur.280...24D}
{Dicke}, R.~H. 1979, \nat, 280, 24, \dodoi{10.1038/280024a0}

\bibitem[{{Esser} \& {Edgar}(2001)}]{2001ApJ...563.1055E}
{Esser}, R., \& {Edgar}, R.~J. 2001, \apj, 563, 1055, \dodoi{10.1086/323987}

\bibitem[{{Esser} {et~al.}(1998){Esser}, {Edgar}, \&
  {Brickhouse}}]{1998ApJ...498..448E}
{Esser}, R., {Edgar}, R.~J., \& {Brickhouse}, N.~S. 1998, \apj, 498, 448,
  \dodoi{10.1086/305516}

\bibitem[{{Fr{\"o}hlich}(2003)}]{2003ESASP.535..183F}
{Fr{\"o}hlich}, C. 2003, in ESA Special Publication, Vol. 535, Solar
  Variability as an Input to the Earth's Environment, ed. A.~{Wilson}, 183--193

\bibitem[{{Fr{\"o}hlich}(2013)}]{2013SSRv..176..237F}
{Fr{\"o}hlich}, C. 2013, \ssr, 176, 237, \dodoi{10.1007/s11214-011-9780-1}

\bibitem[{{Fu} {et~al.}(2015){Fu}, {Li}, {Li}, {Huang}, {Mou}, {Jiao}, \&
  {Xia}}]{2015SoPh..290.1399F}
{Fu}, H., {Li}, B., {Li}, X., {et~al.} 2015, \solphys, 290, 1399,
  \dodoi{10.1007/s11207-015-0689-9}

\bibitem[{{Fu} {et~al.}(2018){Fu}, {Madjarska}, {Li}, {Xia}, \&
  {Huang}}]{2018MNRAS.478.1884F}
{Fu}, H., {Madjarska}, M.~S., {Li}, B., {Xia}, L., \& {Huang}, Z. 2018, \mnras,
  478, 1884, \dodoi{10.1093/mnras/sty1211}

\bibitem[{{Fu} {et~al.}(2017){Fu}, {Madjarska}, {Xia}, {Li}, {Huang}, \&
  {Wangguan}}]{2017ApJ...836..169F}
{Fu}, H., {Madjarska}, M.~S., {Xia}, L., {et~al.} 2017, \apj, 836, 169,
  \dodoi{10.3847/1538-4357/aa5cba}

\bibitem[{{Gazis}(1996)}]{1996RvGeo..34..379G}
{Gazis}, P.~R. 1996, Reviews of Geophysics, 34, 379, \dodoi{10.1029/96RG00892}

\bibitem[{{Gibson} {et~al.}(2009){Gibson}, {Kozyra}, {de Toma}, {Emery},
  {Onsager}, \& {Thompson}}]{2009JGRA..114.9105G}
{Gibson}, S.~E., {Kozyra}, J.~U., {de Toma}, G., {et~al.} 2009, Journal of
  Geophysical Research (Space Physics), 114, A09105,
  \dodoi{10.1029/2009JA014342}

\bibitem[{{Gloeckler} {et~al.}(1998){Gloeckler}, {Cain}, {Ipavich}, {Tums},
  {Bedini}, {Fisk}, {Zurbuchen}, {Bochsler}, {Fischer}, {Wimmer-Schweingruber},
  {Geiss}, \& {Kallenbach}}]{1998SSRv...86..497G}
{Gloeckler}, G., {Cain}, J., {Ipavich}, F.~M., {et~al.} 1998, \ssr, 86, 497,
  \dodoi{10.1023/A:1005036131689}

\bibitem[{{Goelzer} {et~al.}(2013){Goelzer}, {Smith}, {Schwadron}, \&
  {McCracken}}]{2013JGRA..118.7525G}
{Goelzer}, M.~L., {Smith}, C.~W., {Schwadron}, N.~A., \& {McCracken}, K.~G.
  2013, Journal of Geophysical Research (Space Physics), 118, 7525,
  \dodoi{10.1002/2013JA019404}

\bibitem[{{Gopalswamy} {et~al.}(2020){Gopalswamy}, {Akiyama}, {Yashiro},
  {Michalek}, {Xie}, \& {M{\"a}kel{\"a}}}]{2020JPhCS1620a2005G}
{Gopalswamy}, N., {Akiyama}, S., {Yashiro}, S., {et~al.} 2020, in Journal of
  Physics Conference Series, Vol. 1620, Journal of Physics Conference Series,
  012005, \dodoi{10.1088/1742-6596/1620/1/012005}

\bibitem[{{Habbal} {et~al.}(2013){Habbal}, {Morgan}, {Druckm{\"u}ller}, {Ding},
  {Cooper}, {Daw}, \& {Sittler}}]{2013SoPh..285....9H}
{Habbal}, S.~R., {Morgan}, H., {Druckm{\"u}ller}, M., {et~al.} 2013, \solphys,
  285, 9, \dodoi{10.1007/s11207-012-0115-5}

\bibitem[{{Habbal} {et~al.}(2010){Habbal}, {Druckm{\"u}ller}, {Morgan}, {Daw},
  {Johnson}, {Ding}, {Arndt}, {Esser}, {Ru{\v{s}}in}, \&
  {Scholl}}]{2010ApJ...708.1650H}
{Habbal}, S.~R., {Druckm{\"u}ller}, M., {Morgan}, H., {et~al.} 2010, \apj, 708,
  1650, \dodoi{10.1088/0004-637X/708/2/1650}

\bibitem[{{Habbal} {et~al.}(2021){Habbal}, {Druckm{\"u}ller}, {Alzate}, {Ding},
  {Johnson}, {Starha}, {Hoderova}, {Boe}, {Constantinou}, \&
  {Arndt}}]{2021ApJ...911L...4H}
{Habbal}, S.~R., {Druckm{\"u}ller}, M., {Alzate}, N., {et~al.} 2021, \apjl,
  911, L4, \dodoi{10.3847/2041-8213/abe775}

\bibitem[{{Hathaway}(2015)}]{2015LRSP...12....4H}
{Hathaway}, D.~H. 2015, Living Reviews in Solar Physics, 12, 4,
  \dodoi{10.1007/lrsp-2015-4}

\bibitem[{{Heath} \& {Schlesinger}(1986)}]{1986JGR....91.8672H}
{Heath}, D.~F., \& {Schlesinger}, B.~M. 1986, \jgr, 91, 8672,
  \dodoi{10.1029/JD091iD08p08672}

\bibitem[{{Hefti} {et~al.}(2000){Hefti}, {Gr{\"u}nwaldt}, {Bochsler}, \&
  {Aellig}}]{2000JGR...10510527H}
{Hefti}, S., {Gr{\"u}nwaldt}, H., {Bochsler}, P., \& {Aellig}, M.~R. 2000,
  \jgr, 105, 10527, \dodoi{10.1029/1999JA900384}

\bibitem[{{Higginson} {et~al.}(2017){Higginson}, {Antiochos}, {DeVore},
  {Wyper}, \& {Zurbuchen}}]{2017ApJ...840L..10H}
{Higginson}, A.~K., {Antiochos}, S.~K., {DeVore}, C.~R., {Wyper}, P.~F., \&
  {Zurbuchen}, T.~H. 2017, \apjl, 840, L10, \dodoi{10.3847/2041-8213/aa6d72}

\bibitem[{{Issautier} {et~al.}(2008){Issautier}, {Le Chat}, {Meyer-Vernet},
  {Moncuquet}, {Hoang}, {MacDowall}, \& {McComas}}]{2008GeoRL..3519101I}
{Issautier}, K., {Le Chat}, G., {Meyer-Vernet}, N., {et~al.} 2008, \grl, 35,
  L19101, \dodoi{10.1029/2008GL034912}

\bibitem[{{Kasper} {et~al.}(2012){Kasper}, {Stevens}, {Korreck}, {Maruca},
  {Kiefer}, {Schwadron}, \& {Lepri}}]{2012ApJ...745..162K}
{Kasper}, J.~C., {Stevens}, M.~L., {Korreck}, K.~E., {et~al.} 2012, \apj, 745,
  162, \dodoi{10.1088/0004-637X/745/2/162}

\bibitem[{{Kasper} {et~al.}(2007){Kasper}, {Stevens}, {Lazarus}, {Steinberg},
  \& {Ogilvie}}]{2007ApJ...660..901K}
{Kasper}, J.~C., {Stevens}, M.~L., {Lazarus}, A.~J., {Steinberg}, J.~T., \&
  {Ogilvie}, K.~W. 2007, \apj, 660, 901, \dodoi{10.1086/510842}

\bibitem[{{Kirk} {et~al.}(2009){Kirk}, {Pesnell}, {Young}, \& {Hess
  Webber}}]{2009SoPh..257...99K}
{Kirk}, M.~S., {Pesnell}, W.~D., {Young}, C.~A., \& {Hess Webber}, S.~A. 2009,
  \solphys, 257, 99, \dodoi{10.1007/s11207-009-9369-y}

\bibitem[{{Kleczek}(1952)}]{1952BAICz...3...52K}
{Kleczek}, J. 1952, Bulletin of the Astronomical Institutes of Czechoslovakia,
  3, 52

\bibitem[{{Ko} {et~al.}(1997){Ko}, {Fisk}, {Geiss}, {Gloeckler}, \&
  {Guhathakurta}}]{1997SoPh..171..345K}
{Ko}, Y.-K., {Fisk}, L.~A., {Geiss}, J., {Gloeckler}, G., \& {Guhathakurta}, M.
  1997, \solphys, 171, 345

\bibitem[{{Landi} {et~al.}(2012){Landi}, {Gruesbeck}, {Lepri}, {Zurbuchen}, \&
  {Fisk}}]{2012ApJ...758L..21L}
{Landi}, E., {Gruesbeck}, J.~R., {Lepri}, S.~T., {Zurbuchen}, T.~H., \& {Fisk},
  L.~A. 2012, \apjl, 758, L21, \dodoi{10.1088/2041-8205/758/1/L21}

\bibitem[{{Lee} {et~al.}(2009){Lee}, {Luhmann}, {Zhao}, {Liu}, {Riley}, {Arge},
  {Russell}, \& {de Pater}}]{2009SoPh..256..345L}
{Lee}, C.~O., {Luhmann}, J.~G., {Zhao}, X.~P., {et~al.} 2009, \solphys, 256,
  345, \dodoi{10.1007/s11207-009-9345-6}

\bibitem[{{Lemen} {et~al.}(2012){Lemen}, {Title}, {Akin}, {Boerner}, {Chou},
  {Drake}, {Duncan}, {Edwards}, {Friedlaender}, {Heyman}, {Hurlburt}, {Katz},
  {Kushner}, {Levay}, {Lindgren}, {Mathur}, {McFeaters}, {Mitchell}, {Rehse},
  {Schrijver}, {Springer}, {Stern}, {Tarbell}, {Wuelser}, {Wolfson}, {Yanari},
  {Bookbinder}, {Cheimets}, {Caldwell}, {Deluca}, {Gates}, {Golub}, {Park},
  {Podgorski}, {Bush}, {Scherrer}, {Gummin}, {Smith}, {Auker}, {Jerram},
  {Pool}, {Soufli}, {Windt}, {Beardsley}, {Clapp}, {Lang}, \&
  {Waltham}}]{2012SoPh..275...17L}
{Lemen}, J.~R., {Title}, A.~M., {Akin}, D.~J., {et~al.} 2012, \solphys, 275,
  17, \dodoi{10.1007/s11207-011-9776-8}

\bibitem[{{Lepping} {et~al.}(1995){Lepping}, {Ac{\~{u}}na}, {Burlaga},
  {Farrell}, {Slavin}, {Schatten}, {Mariani}, {Ness}, {Neubauer}, {Whang},
  {Byrnes}, {Kennon}, {Panetta}, {Scheifele}, \&
  {Worley}}]{1995SSRv...71..207L}
{Lepping}, R.~P., {Ac{\~{u}}na}, M.~H., {Burlaga}, L.~F., {et~al.} 1995, \ssr,
  71, 207, \dodoi{10.1007/BF00751330}

\bibitem[{{Liewer} {et~al.}(2004){Liewer}, {Neugebauer}, \&
  {Zurbuchen}}]{2004SoPh..223..209L}
{Liewer}, P.~C., {Neugebauer}, M., \& {Zurbuchen}, T. 2004, \solphys, 223, 209,
  \dodoi{10.1007/s11207-004-1105-z}

\bibitem[{{Lockwood}(2003)}]{2003JGRA..108.1128L}
{Lockwood}, M. 2003, Journal of Geophysical Research (Space Physics), 108,
  1128, \dodoi{10.1029/2002JA009431}

\bibitem[{{Lowder} {et~al.}(2017){Lowder}, {Qiu}, \&
  {Leamon}}]{2017SoPh..292...18L}
{Lowder}, C., {Qiu}, J., \& {Leamon}, R. 2017, \solphys, 292, 18,
  \dodoi{10.1007/s11207-016-1041-8}

\bibitem[{{McComas} {et~al.}(2008){McComas}, {Ebert}, {Elliott}, {Goldstein},
  {Gosling}, {Schwadron}, \& {Skoug}}]{2008GeoRL..3518103M}
{McComas}, D.~J., {Ebert}, R.~W., {Elliott}, H.~A., {et~al.} 2008, \grl, 35,
  L18103, \dodoi{10.1029/2008GL034896}

\bibitem[{{Neugebauer} {et~al.}(1998){Neugebauer}, {Forsyth}, {Galvin},
  {Harvey}, {Hoeksema}, {Lazarus}, {Lepping}, {Linker}, {Mikic}, {Steinberg},
  {von Steiger}, {Wang}, \& {Wimmer-Schweingruber}}]{1998JGR...10314587N}
{Neugebauer}, M., {Forsyth}, R.~J., {Galvin}, A.~B., {et~al.} 1998, \jgr, 103,
  14587, \dodoi{10.1029/98JA00798}

\bibitem[{{Ogilvie} {et~al.}(1995){Ogilvie}, {Chornay}, {Fritzenreiter},
  {Hunsaker}, {Keller}, {Lobell}, {Miller}, {Scudder}, {Sittler}, {Torbert},
  {Bodet}, {Needell}, {Lazarus}, {Steinberg}, {Tappan}, {Mavretic}, \&
  {Gergin}}]{1995SSRv...71...55O}
{Ogilvie}, K.~W., {Chornay}, D.~J., {Fritzenreiter}, R.~J., {et~al.} 1995,
  \ssr, 71, 55, \dodoi{10.1007/BF00751326}

\bibitem[{{Owens} {et~al.}(2008){Owens}, {Crooker}, {Schwadron}, {Horbury},
  {Yashiro}, {Xie}, {St. Cyr}, \& {Gopalswamy}}]{2008GeoRL..3520108O}
{Owens}, M.~J., {Crooker}, N.~U., {Schwadron}, N.~A., {et~al.} 2008, \grl, 35,
  L20108, \dodoi{10.1029/2008GL035813}

\bibitem[{{Owocki} {et~al.}(1983){Owocki}, {Holzer}, \&
  {Hundhausen}}]{1983ApJ...275..354O}
{Owocki}, S.~P., {Holzer}, T.~E., \& {Hundhausen}, A.~J. 1983, \apj, 275, 354,
  \dodoi{10.1086/161538}

\bibitem[{{{\"O}zg{\"u}{\c{c}}} {et~al.}(2003){{\"O}zg{\"u}{\c{c}}},
  {Ata{\c{c}}}, \& {Ryb{\'a}k}}]{2003SoPh..214..375O}
{{\"O}zg{\"u}{\c{c}}}, A., {Ata{\c{c}}}, T., \& {Ryb{\'a}k}, J. 2003, \solphys,
  214, 375, \dodoi{10.1023/A:1024225802080}

\bibitem[{{{\"O}zg{\"u}{\c{c}}} {et~al.}(2004){{\"O}zg{\"u}{\c{c}}},
  {Ata{\c{c}}}, \& {Ryb{\'a}k}}]{2004SoPh..223..287O}
---. 2004, \solphys, 223, 287, \dodoi{10.1007/s11207-004-7304-9}

\bibitem[{{Rahmanifard} {et~al.}(2017){Rahmanifard}, {Schwadron}, {Smith},
  {McCracken}, {Duderstadt}, {Lugaz}, \& {Goelzer}}]{2017ApJ...837..165R}
{Rahmanifard}, F., {Schwadron}, N.~A., {Smith}, C.~W., {et~al.} 2017, \apj,
  837, 165, \dodoi{10.3847/1538-4357/aa6191}

\bibitem[{{Ramesh} \& {Rohini}(2008)}]{2008ApJ...686L..41R}
{Ramesh}, K.~B., \& {Rohini}, V.~S. 2008, \apjl, 686, L41,
  \dodoi{10.1086/592774}

\bibitem[{{Ramesh} \& {Vasantharaju}(2014)}]{2014Ap&SS.350..479R}
{Ramesh}, K.~B., \& {Vasantharaju}, N. 2014, \apss, 350, 479,
  \dodoi{10.1007/s10509-014-1804-3}

\bibitem[{{Richardson} \& {Cane}(2004)}]{2004JGRA..109.9104R}
{Richardson}, I.~G., \& {Cane}, H.~V. 2004, Journal of Geophysical Research
  (Space Physics), 109, A09104, \dodoi{10.1029/2004JA010598}

\bibitem[{{Richardson} \& {Cane}(2010)}]{2010SoPh..264..189R}
---. 2010, \solphys, 264, 189, \dodoi{10.1007/s11207-010-9568-6}

\bibitem[{{Rybansk{\'y}} {et~al.}(2005){Rybansk{\'y}}, {Ru{\v{s}}in},
  {Minarovjech}, {Klocok}, \& {Cliver}}]{2005JGRA..110.8106R}
{Rybansk{\'y}}, M., {Ru{\v{s}}in}, V., {Minarovjech}, M., {Klocok}, L., \&
  {Cliver}, E.~W. 2005, Journal of Geophysical Research (Space Physics), 110,
  A08106, \dodoi{10.1029/2005JA011146}

\bibitem[{{Schatten} {et~al.}(1969){Schatten}, {Wilcox}, \&
  {Ness}}]{1969SoPh....6..442S}
{Schatten}, K.~H., {Wilcox}, J.~M., \& {Ness}, N.~F. 1969, \solphys, 6, 442,
  \dodoi{10.1007/BF00146478}

\bibitem[{{Scherrer} {et~al.}(1995){Scherrer}, {Bogart}, {Bush}, {Hoeksema},
  {Kosovichev}, {Schou}, {Rosenberg}, {Springer}, {Tarbell}, {Title},
  {Wolfson}, {Zayer}, \& {MDI Engineering Team}}]{1995SoPh..162..129S}
{Scherrer}, P.~H., {Bogart}, R.~S., {Bush}, R.~I., {et~al.} 1995, \solphys,
  162, 129, \dodoi{10.1007/BF00733429}

\bibitem[{{Schou} {et~al.}(2012){Schou}, {Scherrer}, {Bush}, {Wachter},
  {Couvidat}, {Rabello-Soares}, {Bogart}, {Hoeksema}, {Liu}, {Duvall}, {Akin},
  {Allard}, {Miles}, {Rairden}, {Shine}, {Tarbell}, {Title}, {Wolfson},
  {Elmore}, {Norton}, \& {Tomczyk}}]{2012SoPh..275..229S}
{Schou}, J., {Scherrer}, P.~H., {Bush}, R.~I., {et~al.} 2012, \solphys, 275,
  229, \dodoi{10.1007/s11207-011-9842-2}

\bibitem[{{Schrijver} \& {De Rosa}(2003)}]{2003SoPh..212..165S}
{Schrijver}, C.~J., \& {De Rosa}, M.~L. 2003, \solphys, 212, 165,
  \dodoi{10.1023/A:1022908504100}

\bibitem[{{Schwabe}(1844)}]{1844AN.....21..233S}
{Schwabe}, H. 1844, Astronomische Nachrichten, 21, 233,
  \dodoi{10.1002/asna.18440211505}

\bibitem[{{Schwenn}(2006)}]{2006SSRv..124...51S}
{Schwenn}, R. 2006, \ssr, 124, 51, \dodoi{10.1007/s11214-006-9099-5}

\bibitem[{{Sheeley} {et~al.}(1997){Sheeley}, {Wang}, {Hawley}, {Brueckner},
  {Dere}, {Howard}, {Koomen}, {Korendyke}, {Michels}, {Paswaters}, {Socker},
  {St. Cyr}, {Wang}, {Lamy}, {Llebaria}, {Schwenn}, {Simnett}, {Plunkett}, \&
  {Biesecker}}]{1997ApJ...484..472S}
{Sheeley}, N.~R., {Wang}, Y.~M., {Hawley}, S.~H., {et~al.} 1997, \apj, 484,
  472, \dodoi{10.1086/304338}

\bibitem[{{Shi} {et~al.}(2022){Shi}, {Fu}, {Huang}, {Ma}, \&
  {Xia}}]{2022ApJ...940..103S}
{Shi}, X., {Fu}, H., {Huang}, Z., {Ma}, C., \& {Xia}, L. 2022, \apj, 940, 103,
  \dodoi{10.3847/1538-4357/ac9b20}

\bibitem[{{Smith} \& {Balogh}(2008)}]{2008GeoRL..3522103S}
{Smith}, E.~J., \& {Balogh}, A. 2008, \grl, 35, L22103,
  \dodoi{10.1029/2008GL035345}

\bibitem[{{Solanki} {et~al.}(2005){Solanki}, {Krivova}, \&
  {Wenzler}}]{2005AdSpR..35..376S}
{Solanki}, S.~K., {Krivova}, N.~A., \& {Wenzler}, T. 2005, Advances in Space
  Research, 35, 376, \dodoi{10.1016/j.asr.2004.12.077}

\bibitem[{{Solanki} {et~al.}(2000){Solanki}, {Sch{\"u}ssler}, \&
  {Fligge}}]{2000Natur.408..445S}
{Solanki}, S.~K., {Sch{\"u}ssler}, M., \& {Fligge}, M. 2000, \nat, 408, 445,
  \dodoi{10.1038/35044027}

\bibitem[{{Song} {et~al.}(2022){Song}, {Cheng}, {Li}, {Zhang}, \&
  {Chen}}]{2022ApJ...925..137S}
{Song}, H., {Cheng}, X., {Li}, L., {Zhang}, J., \& {Chen}, Y. 2022, \apj, 925,
  137, \dodoi{10.3847/1538-4357/ac3bbf}

\bibitem[{{Song} {et~al.}(2009){Song}, {Chen}, {Liu}, {Feng}, \&
  {Xia}}]{2009SoPh..258..129S}
{Song}, H.~Q., {Chen}, Y., {Liu}, K., {Feng}, S.~W., \& {Xia}, L.~D. 2009,
  \solphys, 258, 129, \dodoi{10.1007/s11207-009-9411-0}

\bibitem[{{Song} {et~al.}(2012){Song}, {Kong}, {Chen}, {Li}, {Li}, {Feng}, \&
  {Xia}}]{2012SoPh..276..261S}
{Song}, H.~Q., {Kong}, X.~L., {Chen}, Y., {et~al.} 2012, \solphys, 276, 261,
  \dodoi{10.1007/s11207-011-9848-9}

\bibitem[{{Stone} {et~al.}(1998){Stone}, {Frandsen}, {Mewaldt}, {Christian},
  {Margolies}, {Ormes}, \& {Snow}}]{1998SSRv...86....1S}
{Stone}, E.~C., {Frandsen}, A.~M., {Mewaldt}, R.~A., {et~al.} 1998, \ssr, 86,
  1, \dodoi{10.1023/A:1005082526237}

\bibitem[{{Suess} {et~al.}(2009){Suess}, {Ko}, {von Steiger}, \&
  {Moore}}]{2009JGRA..114.4103S}
{Suess}, S.~T., {Ko}, Y.~K., {von Steiger}, R., \& {Moore}, R.~L. 2009, Journal
  of Geophysical Research (Space Physics), 114, A04103,
  \dodoi{10.1029/2008JA013704}

\bibitem[{{Temmer} {et~al.}(2002){Temmer}, {Veronig}, \&
  {Hanslmeier}}]{2002A&A...390..707T}
{Temmer}, M., {Veronig}, A., \& {Hanslmeier}, A. 2002, \aap, 390, 707,
  \dodoi{10.1051/0004-6361:20020758}

\bibitem[{{Temmer} {et~al.}(2003){Temmer}, {Veronig}, \&
  {Hanslmeier}}]{2003SoPh..215..111T}
---. 2003, \solphys, 215, 111, \dodoi{10.1023/A:1024843010048}

\bibitem[{{Wang}(2009)}]{2009SSRv..144..383W}
{Wang}, Y.~M. 2009, \ssr, 144, 383, \dodoi{10.1007/s11214-008-9434-0}

\bibitem[{{Wang} {et~al.}(2009){Wang}, {Ko}, \&
  {Grappin}}]{2009ApJ...691..760W}
{Wang}, Y.~M., {Ko}, Y.~K., \& {Grappin}, R. 2009, \apj, 691, 760,
  \dodoi{10.1088/0004-637X/691/1/760}

\bibitem[{{Wang} \& {Sheeley}(1990)}]{1990ApJ...355..726W}
{Wang}, Y.~M., \& {Sheeley}, N.~R., J. 1990, \apj, 355, 726,
  \dodoi{10.1086/168805}

\bibitem[{{Wang} {et~al.}(1997){Wang}, {Sheeley}, {Phillips}, \&
  {Goldstein}}]{1997ApJ...488L..51W}
{Wang}, Y.~M., {Sheeley}, N.~R., J., {Phillips}, J.~L., \& {Goldstein}, B.~E.
  1997, \apjl, 488, L51, \dodoi{10.1086/310918}

\bibitem[{{Woods} {et~al.}(2000){Woods}, {Tobiska}, {Rottman}, \&
  {Worden}}]{2000JGR...10527195W}
{Woods}, T.~N., {Tobiska}, W.~K., {Rottman}, G.~J., \& {Worden}, J.~R. 2000,
  \jgr, 105, 27195, \dodoi{10.1029/2000JA000051}

\bibitem[{{Zerbo} \& {Richardson}(2015)}]{2015JGRA..12010250Z}
{Zerbo}, J.~L., \& {Richardson}, J.~D. 2015, Journal of Geophysical Research
  (Space Physics), 120, 10,250, \dodoi{10.1002/2015JA021407}

\bibitem[{{Zhao} {et~al.}(2017){Zhao}, {Landi}, {Lepri}, {Gilbert},
  {Zurbuchen}, {Fisk}, \& {Raines}}]{2017ApJ...846..135Z}
{Zhao}, L., {Landi}, E., {Lepri}, S.~T., {et~al.} 2017, \apj, 846, 135,
  \dodoi{10.3847/1538-4357/aa850c}

\bibitem[{{Zhao} {et~al.}(2009){Zhao}, {Zurbuchen}, \&
  {Fisk}}]{2009GeoRL..3614104Z}
{Zhao}, L., {Zurbuchen}, T.~H., \& {Fisk}, L.~A. 2009, \grl, 36, L14104,
  \dodoi{10.1029/2009GL039181}

\end{thebibliography}
\bibliographystyle{aasjournal}
\end{document}